\documentclass{sig-alternate-2013}

\usepackage{cite}

\usepackage{subcaption}
\usepackage{graphicx}
\usepackage{amsmath,amssymb,amscd}
\usepackage{soul}
\usepackage{color}
\usepackage{balance}
\usepackage{paralist}
\usepackage{url}

\usepackage[
  pass,
]{geometry}

\newcommand{\pk}{pk^{ABE}}
\newcommand{\mk}{mk^{ABE}}

\newcommand{\andraben}{\textsc{AndrABEn}}

\newfont{\mycrnotice}{ptmr8t at 7pt}
\newfont{\myconfname}{ptmri8t at 7pt}
\let\crnotice\mycrnotice%
\let\confname\myconfname%

\permission{Permission to make digital or hard copies of all or part of this work for personal or classroom use is granted without fee provided that copies are not made or distributed for profit or commercial advantage and that copies bear this notice and the full citation on the first page. Copyrights for components of this work owned by others than ACM must be honored. Abstracting with credit is permitted. To copy otherwise, or republish, to post on servers or to redistribute to lists, requires prior specific permission and/or a fee. Request permissions from Permissions@acm.org.}
\conferenceinfo{IoT-Sys 2015,}{May 18, 2015, Florence, Italy.}
\copyrightetc{Copyright \copyright~2015 ACM \the\acmcopyr}
\crdata{978-1-4503-3502-7/15/05\ ...\$15.00.\\
http://dx.doi.org/10.1145/2753476.2753482}

\begin{document}

\title{On the Feasibility of Attribute-Based Encryption \\ on Smartphone Devices}

\numberofauthors{3} 

\author{
\alignauthor Moreno Ambrosin\\
       \affaddr{University of Padua, Italy}\\
       \email{ambrosin@math.unipd.it}
\alignauthor Mauro Conti \thanks{\small Mauro Conti is supported by a European Marie Curie Fellowship (N. PCIG11-GA-2012-321980). This work is also partially supported by the Italian MIUR PRIN Project TENACE  (N. 20103P34XC), and the University of Padua PRAT 2014 Project on Mobile Malware.}\\
       \affaddr{University of Padua, Italy}\\
       \email{conti@math.unipd.it}  
\alignauthor Tooska Dargahi\\
       \affaddr{West Tehran Branch,}\\
     \affaddr{Islamic Azad University, Iran}\\
       \email{t.dargahi@srbiau.ac.ir}
}

\maketitle
\begin{abstract}

\sloppy
Attribute-Based Encryption (ABE) is a powerful cryptographic tool that allows fine-grained access control over data.
Due to its features, ABE has been adopted in several applications, such as encrypted storage or access control systems.
Recently, researchers argued about the non acceptable performance of ABE when implemented on mobile devices. Indeed, the non feasibility of ABE on mobile devices would hinder the deployment of novel protocols and services--that could instead exploit the full potential of such devices.
However, we believe the conclusion of non usability was driven by a not-very efficient implementation.

In this paper, we want to shine a light on this concern by studying the feasibility of applying ABE on smartphone devices. 
 In particular, we implemented \andraben, an ABE library for Android operating system. Our library is written in the C language and implements two main ABE schemes:
Ciphertext-Policy Attribute-Based Encryption, and Key- Policy Attribute-Based Encryption. 
We also run a thorough set of experimental evaluation for \andraben, and compare it with the current state-of-the-art (considering the same experimental setting). The results confirm the possibility to effectively use ABE on smartphone devices, requiring an acceptable amount of resources in terms of computations and energy consumption.
Since the current state-of-the-art claims the non feasibility of ABE on mobile devices, we believe that our study (together with the \andraben~library that we made available online) is a key result that will pave the way for researchers and developers to design and implement novel protocols and applications for mobile devices.
\end{abstract}

%
\newpage
\section{Introduction} \label{sec:intro}

\sloppy
Attribute-Based Encryption (ABE) is a public key encryption scheme first introduced in 2005 by Sahai and Waters~\cite{sahai2005fuzzy}. In this scheme, both encryption and decryption are based on {\em attributes} (e.g., age, gender, or job position), that can be either related to the private keys of the users, or to the ciphertext. A user can restrict access to a specific piece of data by defining an {\em access policy}. As an example, an access policy can be expressed as a boolean expression such as $(A \wedge B)\vee C$, where $A$, $B$ and $C$ are attributes and the possible values for attributes are implicitly $true$ or $false$.
Researchers proposed two main types of ABE schemes, namely Key-Policy Attribute-Based Encryption (KP-ABE)~\cite{goyal2006attribute} and Ciphertext-Policy Attribute-Based Encryption (CP-ABE)~\cite{bethencourt2007ciphertext}. 

 Compared to the other encryption approaches, ABE presents several advantages~\cite{lee2013survey}. First, it allows the data owner to apply a fine grained access control over data, based on attributes and policies. Second, ABE schemes are scalable and independent of the number of authorized users. Moreover, ABE by construction is resilient against collusion attacks. Finally, while traditional public key infrastructures impose a noticeable communication and storage overhead due to the exchange of cryptographical material, using ABE the data owner can encrypt data by using a set of attributes, without exchanging any certificates or identifying the client~\cite{lee2013survey}.


Several factors influence the performance of ABE in real applications, such as the number of attributes used for defining an access policy, the desired security level, and the capabilities of the underlying device, in terms of available memory and CPU speed. 
Some researchers have already studied the feasibility of using ABE on mobile devices~\cite{ABE_icc_2014}; however, most of the existing studies do not consider all the factors, or actually do not implement ABE on smartphone. In~\cite{ABE_icc_2014}, Wang et al. evaluated the performance of CP-ABE and KP-ABE on laptop and smartphone devices. They implemented both schemes by Java language, and evaluated different metrics. 
The authors concluded that the ABE performance is unacceptable on Android smartphone. However, the usefulness of ABE in mobile applications is evident, and its non feasibility on such devices would be a big obstacle to deployment of new services and to benefit from its advantages. Therefore, obtaining acceptable performance is the biggest challenge for guaranteeing the use of ABE on resource constraint devices.

\newpage
\paragraph{Contribution} The contribution of this paper is a comprehensive and careful study of the feasibility of ABE operations on Android smartphone devices.
In particular, differently from what claimed in~\cite{ABE_icc_2014}, we show that it is possible to achieve reasonable performance for the main ABE operations, even on smartphone devices. 
We provide \andraben, an implementation of CP-ABE~\cite{bethencourt2007ciphertext} and KP-ABE~\cite{goyal2006attribute} as a C library for Android smartphones. We integrated such library into Android using the Android Native Development Kit (NDK) tool. We evaluated our implementation, and compared its performance with the Java-based study and implementation proposed by Wang et al. in their work presented at ICC 2014~\cite{ABE_icc_2014}, considering the same experimental settings\footnote{We re-implemented the proposal in~\cite{ABE_icc_2014} due to the unavailability of the original source code.}.
The results of our thorough evaluation show that the performance of our solution is an order of magnitude higher than the one in~\cite{ABE_icc_2014}. Accordingly, our results prove that applying ABE is indeed feasible on current mobile devices, such as Samsung Galaxy Nexus smartphone---on which we ran our experiments. Finally, we made the \andraben~library freely available~\cite{andraben_impl} for researchers and developers.

\paragraph{Organization} The rest of the paper is organized as follows. In Section \ref{sec:related} we present some related work. In Section \ref{sec:background} we introduce the preliminaries and background on ABE. In Section \ref{sec:our-implementation}, we present \andraben, our proposed implementation for CP-ABE and KP-ABE, and provide an analysis of its performance. We also discuss and compare our results with the ones in~\cite{ABE_icc_2014}. Finally, in Section \ref{sec:conclusion} we draw our conclusions.

\section{Related work}\label{sec:related}
Due to the increasing use of mobile devices, evaluating the performance of cryptographic algorithms on mobile devices is an important issue that has been considered in several research studies~\cite{tillich2004survey, kawahara2006efficient, voyiatzis2011increasing, braga2012portability}. 
As an example, Braga and Nascimento~\cite{braga2012portability} evaluated the feasibility of cryptographic algorithms on Android smartphone devices. They assessed the portability of cryptographic libraries on a Samsung i9100 and measured the performance of applying these libraries on the Android devices; however, in their analysis they did not consider ABE. 

Recently, the concept of ABE has been used in various schemes to deal with data confidentiality, privacy and access control issues. Unfortunately, most of these research studies did not evaluate the actual feasibility of adopting ABE in their proposed approaches. However, few researchers focused on such assessment; we discuss some of them in the following. 

In~\cite{Baden:2009:POS:1594977.1592585}, Baden et al. presented Persona, an Online Social Network, where users are able to hide their personal information. 
They provided privacy by encrypting data with ABE. They implemented their solution on a first generation iPhone device, 
by cross-compiling the {\tt cpabe} library~\cite{cpabe} and its dependencies for the iPhone SDK 2.2.1. 
They obtained an average decryption and encryption time of 254~ms and 926~ms respectively, considering access structures containing one to five attributes. However, the authors did not provide an implementation of ABE on the Android smartphone.



Along the same line of studies, we also proposed a system for efficient
software updates distribution, over untrusted distribution networks in~\cite{Moreno_ESORICS}. We used CP-ABE to guarantee flexible access control.
In this work, we only provided evaluations measured on a laptop device with two 2.4 GHz Intel Core 2 Duo CPUs. Based on our experimental results, the CP-ABE encryption and decryption time for five attributes are 77.47~ms and 32.62~ms, respectively.
We realized that not providing a proper feasibility assessment is a big limitation for this type of proposals. In this paper, we aim to fill this gap and pave the way for developing further solutions assuming the feasibility of ABE for mobile devices.

A study similar to the one we are going to discuss in this paper was carried out in~\cite{ABE_icc_2014}. The authors evaluated the performance of CP-ABE and KP-ABE in terms of execution time, data overhead, energy consumption, CPU and memory usage. They implemented these two ABE schemes using Java on a laptop with a 1.60~GHz Intel Quad-Core i7 2677M CPU and a smartphone runs Android 4.04 with a 1.60~GHz Intel Atom Z2460. The authors stated that applying ABE on Android smartphone devices is not practical with acceptable performance. 
In this paper, we show that this conclusion mostly depends on the specific implementation provided in~\cite{ABE_icc_2014}, and it does not hold in general. 

Finally, in~\cite{green2011outsourcing} Green et al. proposed an alternative approach for efficient ABE decryption. In their solution, a part of the ABE decryption is outsourced to a third party cloud, highly reducing the load on the client device. However, while representing a good option to facilitate the ABE operations on devices with limited resources, this solution requires additional resources compared to in-device decryption, such as a third-party cloud entity, as well as Internet connectivity.

To the best of our knowledge, we are the first that show the reasonable performance of ABE on Android devices, and provide a publicly available implementation for Android smartphone devices~\cite{andraben_impl}. Indeed, existing ABE implementations for Android showed a high computation overhead on mobile platforms~\cite{6654173, ABE_icc_2014}. 
We prepared and cross-compiled two ABE C libraries, to be used on the Android mobile devices, thus proving the feasibility of such schemes on this platform. 
While we leave this as a future work, we expect that \andraben~can be easily extended for other mobile devices and Internet of Things (IoT) devices.

\section{Background on ABE}\label{sec:background}

This section provides the fundamentals of Key-Policy ABE (KP-ABE)~\cite{goyal2006attribute}, and Ciphertext-Policy ABE (CP-ABE)~\cite{bethencourt2007ciphertext}. Both CP-ABE and KP-ABE, are public key schemes.
In a KP-ABE scheme, the data owner encrypts the data specifying a set of attributes. Each user owns a private key $D$ that reflects a specific policy. She will be able to decrypt a ciphertext if and only if the attributes embedded into the ciphertext satisfy the policy in $D$. 
It consists of four functions:

\begin{itemize}
\item{\bf Setup.} It takes as input an implicit security parameter and outputs the public parameter $\pk$, and a master key $\mk$.

\item{\bf Encryption.} It takes as input a message $M$, a set of attributes $\gamma$, and the public parameter $\pk$, and outputs the ciphertext $E$.

\item{\bf KeyGen.} It takes as input an access policy $A$, the master key $\mk$ and the public parameter $\pk$. It outputs a decryption key $D$ reflecting the given policy.
\item{\bf Decryption.} It takes as input the ciphertext $E$ that is encrypted under the set of attributes $\gamma$; the
decryption key $D$, that represents the access policy $A$; and the public parameter $\pk$. It outputs the message $M$ if and only if $\gamma$ ``satisfies'' the access policy $A$.

\end{itemize}

Different from KP-ABE, in the CP-ABE scheme, the data owner encrypts her data enforcing an access policy. Users are provided with private keys representing a set of attributes. Only users having attributes that satisfy an access policy will be able to decrypt the ciphertext.
A CP-ABE scheme provides the following functions:

\begin{itemize}

\item {\bf Setup.} It takes as input an implicit security parameter and outputs the public parameter $\pk$, and a master key $\mk$.
\item {\bf Encryption.} It takes as input a message $M$, an access policy $A$, and the public parameter $\pk$, and outputs the ciphertext $E$.
\item {\bf KeyGen.} It takes as input a set of attributes $\gamma$, the master key $\mk$ and the public parameter $\pk$. It outputs a decryption key $D$ reflecting the given attributes.
\item {\bf Decryption.} It takes as input the ciphertext $E$ that is encrypted under the access policy $A$; the
decryption key $D$ representing a set of attributes $\gamma$; and the public parameter $\pk$. It outputs the message $M$ if and only if $\gamma$ ``satisfies'' the access policy $A$.

\end{itemize}

Similar to other pairing-based schemes, the complexity of CP-ABE and KP-ABE depends on the number of exponentiations and pairing operations performed by each of their algorithms~\cite{li2010data}. 
In the CP-ABE scheme~\cite{bethencourt2007ciphertext}, the efficiency of the {\bf KeyGen} algorithm depends on the number of attributes to be applied to the newly generated key. Herein, the algorithm performs two exponentiations for each attribute. Similarly, the {\bf Encryption} operation requires two exponentiations for each attribute in the specified policy. The same complexity is required also by the {\bf KeyGen} and {\bf Encryption} operations in the KP-ABE scheme~\cite{goyal2006attribute}. However, the efficiency of the {\bf Decryption} function for CP-ABE mainly depends on how the policy enforced on the ciphertext and on the private key used for its decryption. This makes an estimation of the complexity of such operation a non trivial task~\cite{bethencourt2007ciphertext}. The same holds for the KP-ABE {\bf Decryption} operation, which strongly depends on the attributes set and the access policy specified in the ciphertext and the private key, respectively. 


\section{ANDRABEN: Implementation and Analysis}\label{sec:our-implementation}
In this section, we provide an in-depth analysis of the performance of \andraben~(Section~\ref{sec:performance}), and a comparison with the implementation in~\cite{ABE_icc_2014} (Section~\ref{sec:discussion}).

\subsection{Performance Evaluation}\label{sec:performance}

Our ABE implementation~\cite{andraben_impl} comprises two libraries: the {\tt cpabe} library~\cite{cpabe}, which implements the scheme proposed by Bethencourt et al. in~\cite{bethencourt2007ciphertext}, and a publicly available custom implementation \cite{kpabe_impl} of the KP-ABE scheme proposed by Goyal et al. in~\cite{goyal2006attribute}. 
The original code has been slightly modified, in order to be integrated into Android mobile devices.
Both libraries employ Type $A$ pairings provided by the PBC library~\cite{libpbc}. Type $A$ pairings are built on top of an elliptic curve: $y^2 = x^3 + x$ over a finite field $F_q$, for some prime $q=3~mod~4$, and have a fixed embedding degree $k = 2$ \cite{pbc-thesis}. Therefore, the security strength of the scheme can be tuned by modifying two parameters: the size of the field $q$, and the prime order $r$ of the base point $P\in E(F_q)$~\cite{ABE_icc_2014}.
Table~\ref{tbl:table_strength} shows the security level of both CP-ABE and KP-ABE schemes, according to~\cite{brown2001software}.

\def\arraystretch{1.2}
\begin{table}[h]
	\centering
	\small
	\begin{tabular}{| c | c | c | c |}
		\hline 
		{\bf Security level bits} & \bf 80 & \bf 112 & \bf128 \\ \hline
		bit length of $r$ & 160 & 224 & 256 \\ \hline
		bit length of $q$ & 512 & 1024 & 1536 \\ \hline
	\end{tabular}
	\caption{Security strength of CP-ABE and KP-ABE.}
	\label{tbl:table_strength}
\end{table}

\andraben~is implemented on Android 4.3, ``Jelly Beam''. We carried out our experimental evaluation on a Samsung Galaxy Nexus device (1.2~GHz dual-core ARM Cortex-A9 CPU, 1~GB RAM). For completeness and comparison, we also tested our libraries on a laptop device (Ubuntu 14.4~LTS, 1.8 GHz 4x Intel Core\texttrademark~i7-4500U CPU, 8~GB RAM).
 
 We evaluated {\bf KeyGen}, {\bf Encryption} and {\bf Decryption} operations, varying the number of attributes adopted from one to 30. We consider this range to be representative enough for a wide range of real world applications of ABE~\cite{ABE_icc_2014}. 

 We tested both ABE schemes with security levels of 80, 112 and 128 bits, and measured average execution time, CPU and memory utilization, and energy consumption on mobile devices.
 Note that, as in other public key schemes, in ABE the actual encryption of the ciphertext is performed by means of a symmetric key, which is in turn encrypted with the public key. Therefore, we evaluate both encryption and decryption operations performed on a symmetric key. 
 This makes our analysis independent from the size of the ciphertext.

\paragraph{Execution Time} Figure~\ref{fig:time_android} presents the average time overhead for {\bf Encryption} and {\bf Decryption} operations for both CP-ABE and KP-ABE schemes. The results are presented for both Android and Laptop devices. They have been obtained as an average of 100 executions for each operation, varying the number of employed attributes and adopting different levels of security. 

 \newcommand{\picsize}{.95\columnwidth}
\newcommand{\picboxsize}{0.49\columnwidth}

\begin{figure}[h!]
	\centering
	\begin{subfigure}[b]{\picboxsize}
		\includegraphics[width=.95\columnwidth]{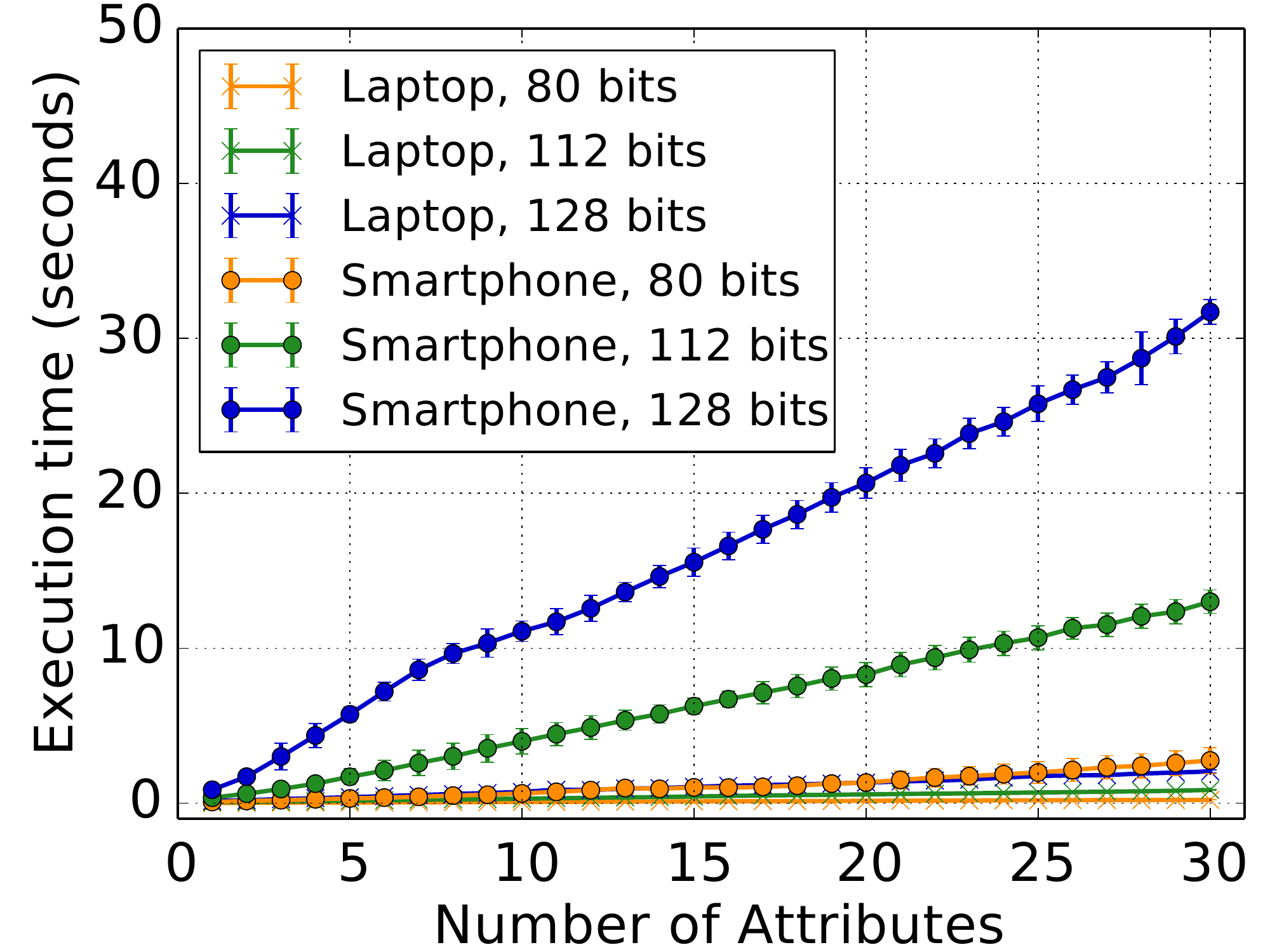}
		\caption{CP-ABE Keygen}
		\label{cpabe_keygen_android}
	\end{subfigure}
	\begin{subfigure}[b]{\picboxsize}
		\includegraphics[width=.95\columnwidth]{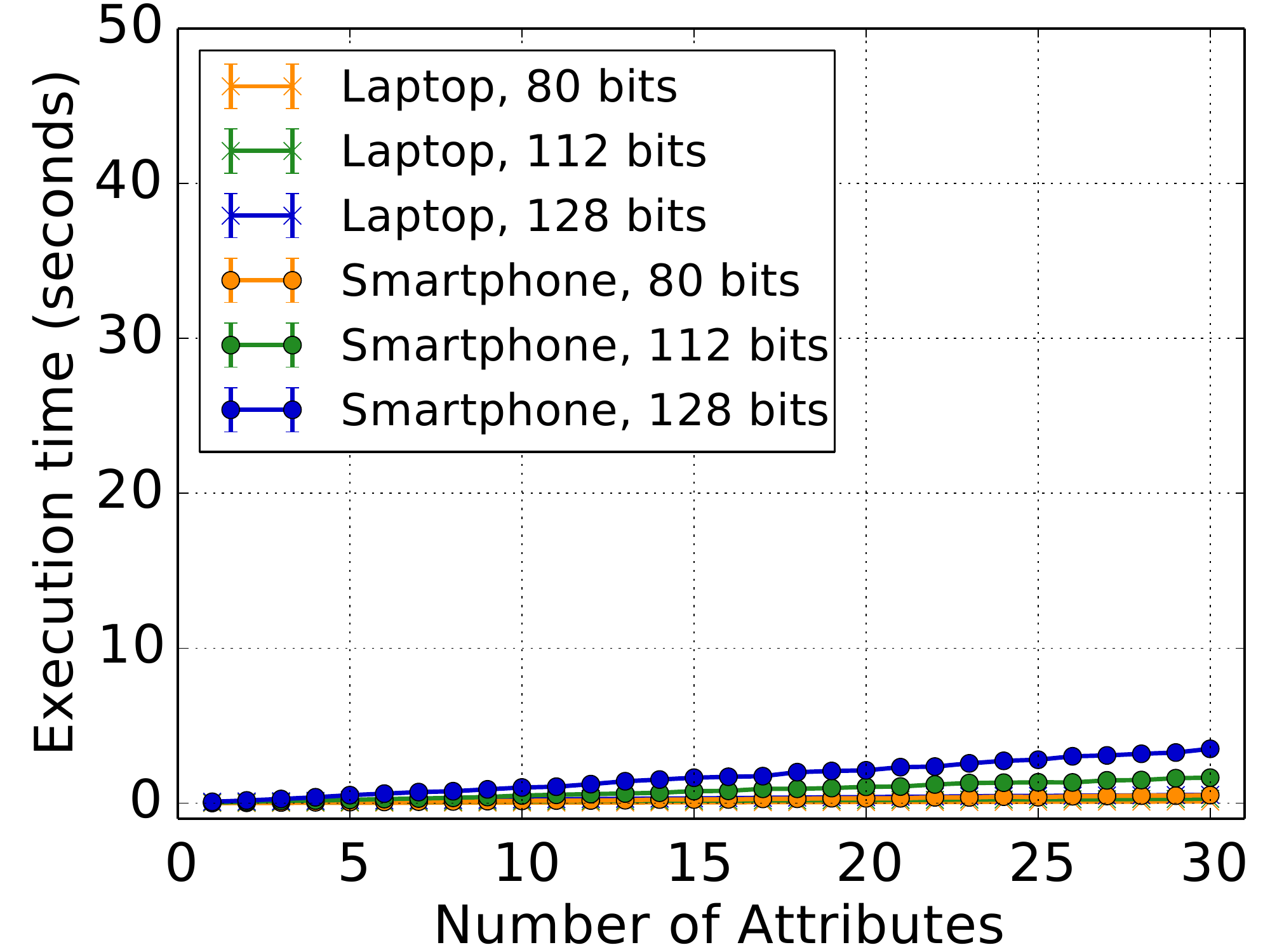}
		\caption{KP-ABE Keygen}
		\label{kpabe_keygen_android}
	\end{subfigure}
	\\
	\begin{subfigure}[b]{\picboxsize}
		\includegraphics[width=.95\columnwidth]{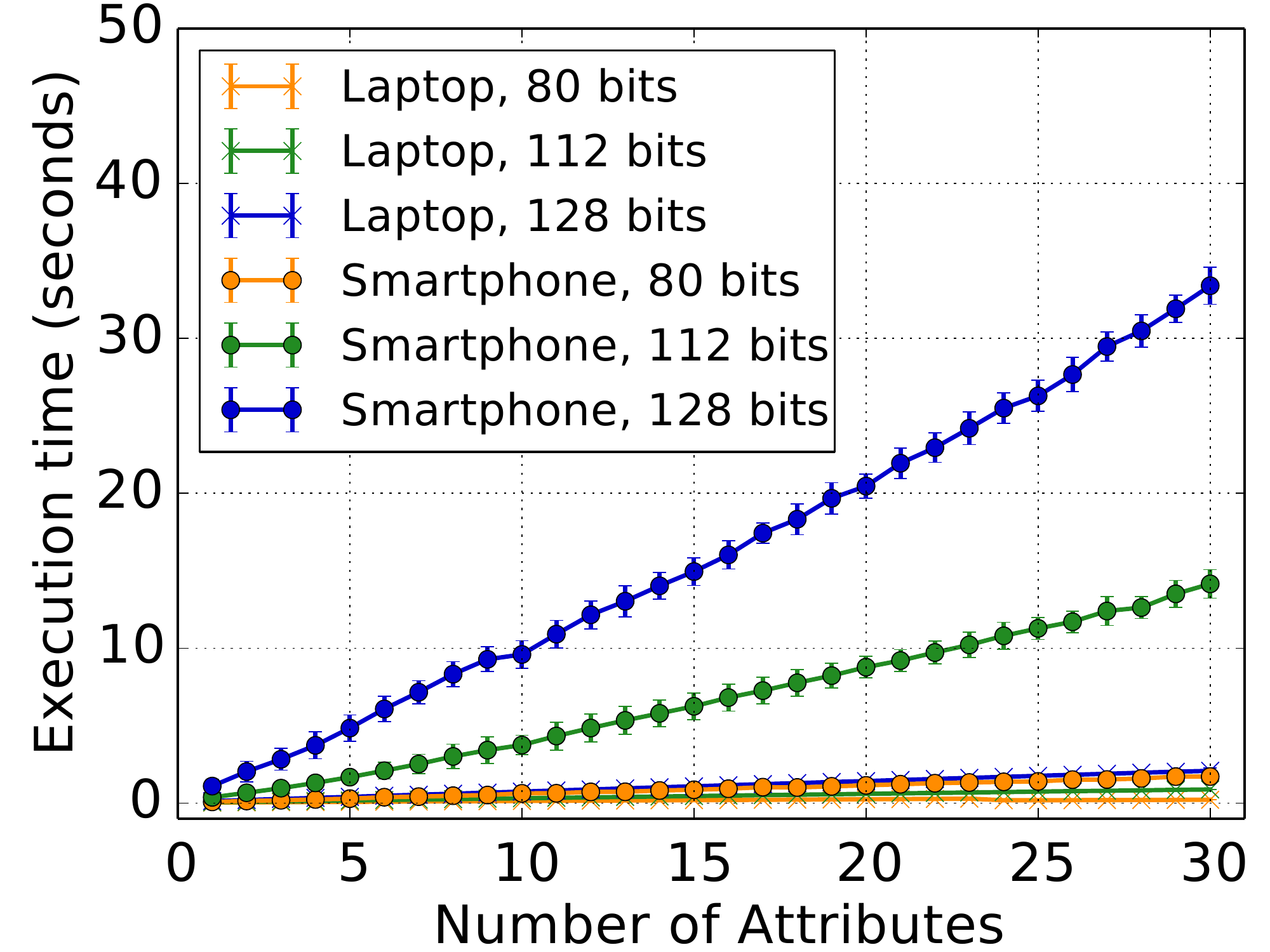}
		\caption{CP-ABE Encryption}
		\label{cpabe_enc_android}
	\end{subfigure}
	\begin{subfigure}[b]{\picboxsize}
		\includegraphics[width=.95\columnwidth]{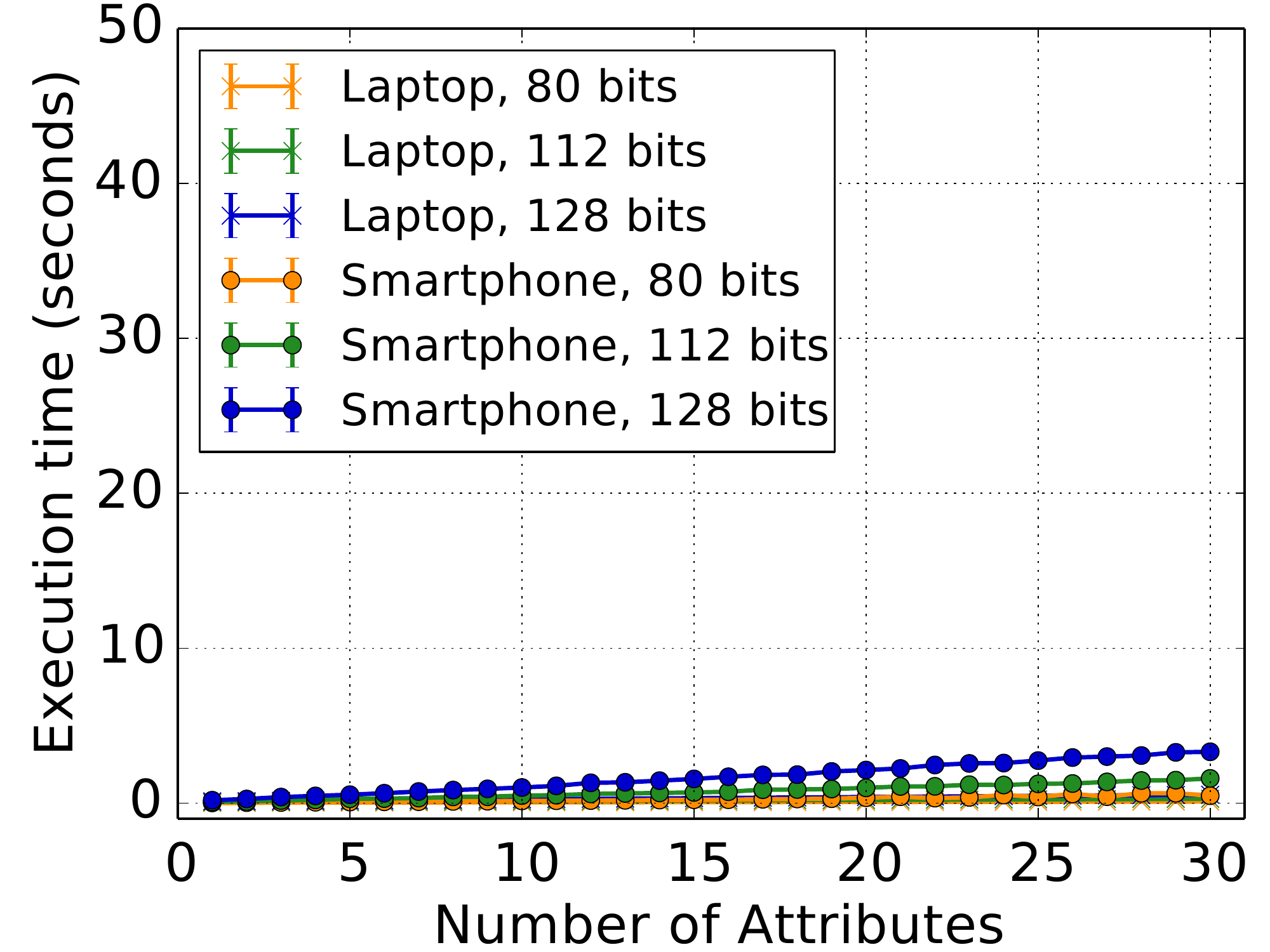}
		\caption{KP-ABE Encryption}
		\label{kpabe_enc_android}
	\end{subfigure}
	\\
	\begin{subfigure}[b]{\picboxsize}
		\includegraphics[width=.95\columnwidth]{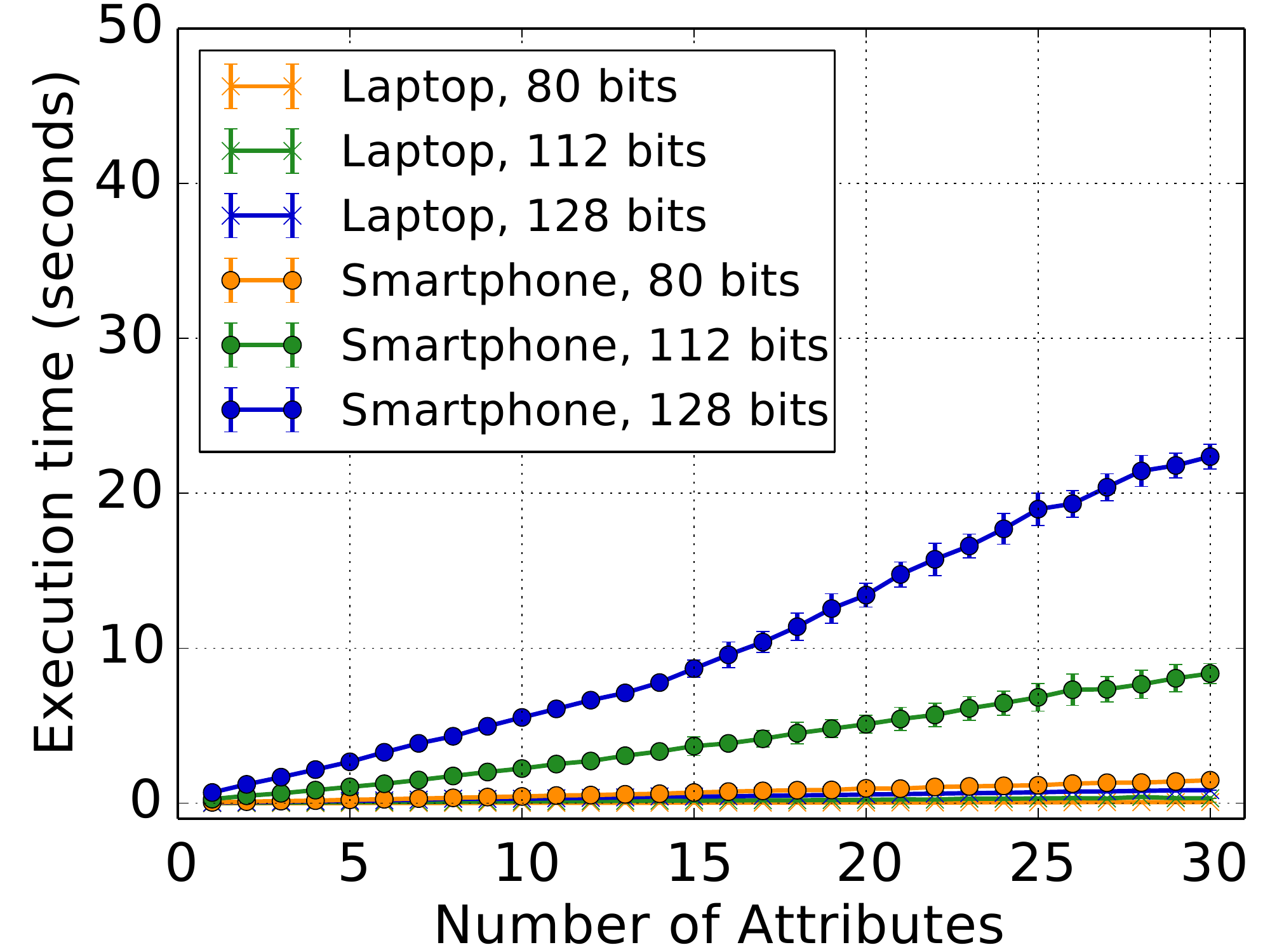}
		\caption{CP-ABE Decryption}
		\label{cpabe_dec_android}
	\end{subfigure}
	\begin{subfigure}[b]{\picboxsize}
		\includegraphics[width=.95\columnwidth]{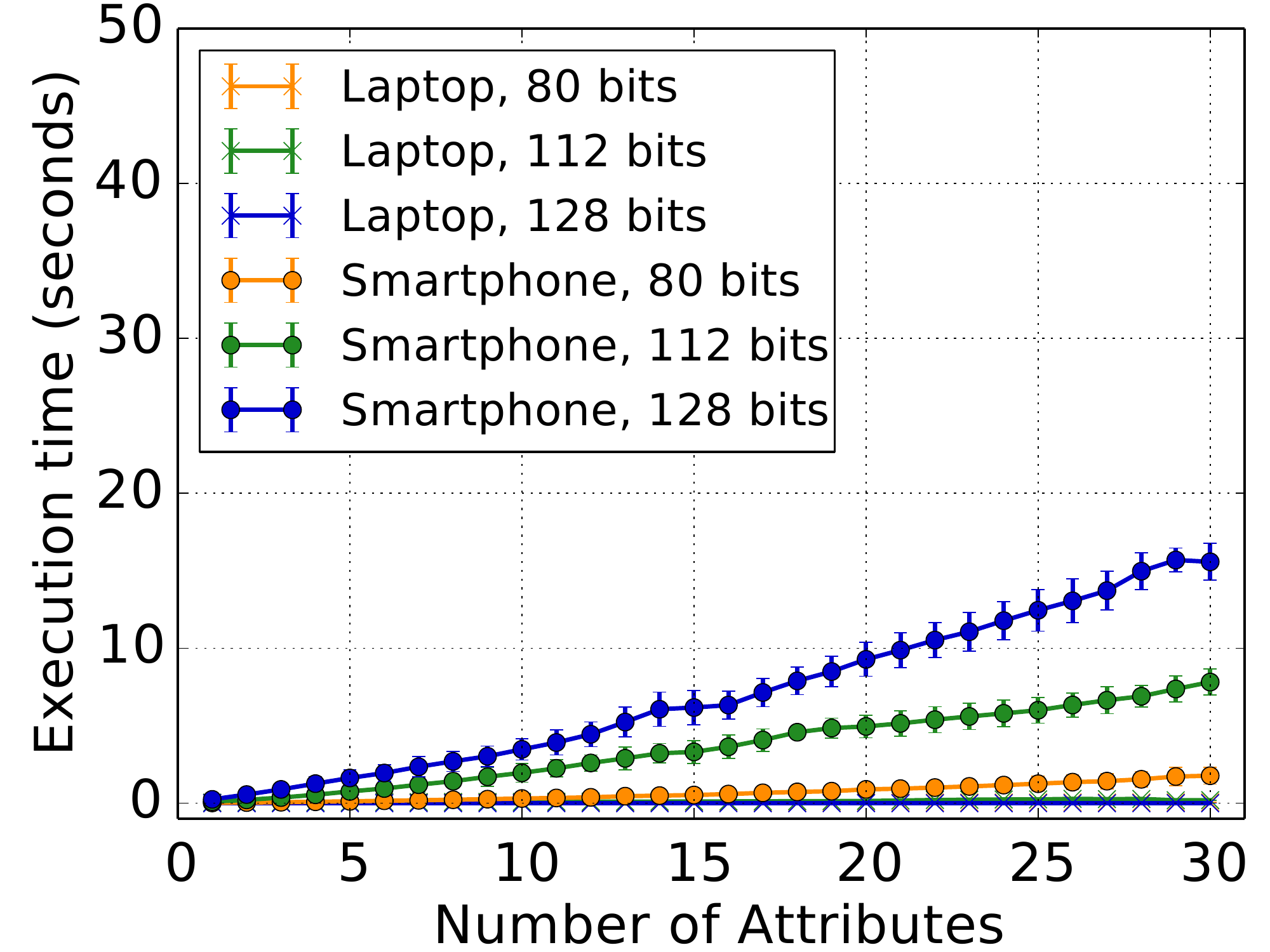}
		\caption{KP-ABE Decryption}
		\label{kpabe_dec_android}
	\end{subfigure}
	\caption{Average execution time (and std. deviation in errorbar) for main CP-ABE and KP-ABE algorithms.}
	\label{fig:time_android}
\end{figure}
 
 As we can see in Figure~\ref{fig:time_android}, in general, the time required to perform each operation depends directly on the number of attributes that are used. Adopting a security value of 80 bits (which is reasonable for several medium-level security applications~\cite{cryptoeprint:2009:389}) the CP-ABE {\bf Encryption} operation remains under 4~s on the Android smartphone. Similarly, the CP-ABE {\bf KeyGen} operation requires less than 2~s to be executed. However, adopting a security level of 112 or 128 bits, the time overhead imposed by CP-ABE on Android smartphone is much higher, while we argue being still usable for non-interactive applications (e.g., encrypted data to be uploaded to a cloud storage service). Indeed, in such applications the encryption could be carried out in a background process. 
 For KP-ABE, however, the time required to perform the various operations is lower, and even the adoption of a security level of 128 bits is feasible on Android smartphone. 
 On laptop, the evaluation results confirm the practicality of both CP-ABE and KP-ABE schemes, requiring a reasonable time for both {\bf KeyGen} ($<\!2$~s), and {\bf Encryption} ($<\!2$~s) operations.
 
\paragraph{CPU utilization} We measured CPU utilization on Android smartphone by collecting the required information from the system files {\tt /proc/stat}, and {\tt /proc/[pid]/stat}, where {\tt pid} is the id of the application's process. The CPU utilization remains under $50\%$ for each of the three operations, for both CP-ABE and KP-ABE, i.e., the operations fully utilize one of the two CPUs provided by the underlying platform.

\paragraph{Memory utilization} We measured the average memory space required by \andraben, adopting a range between one to 30 attributes. We realized that our implementations utilize between $13.5$ and $14.5$ MBytes of RAM space. We argue that such amount is acceptable for modern smartphones such as the Samsung Galaxy Nexus---used in our experiments. 

\begin{figure}[h!]
	\centering
	\begin{subfigure}[b]{0.49\columnwidth}
		\centering
		\includegraphics[width=.95\columnwidth]{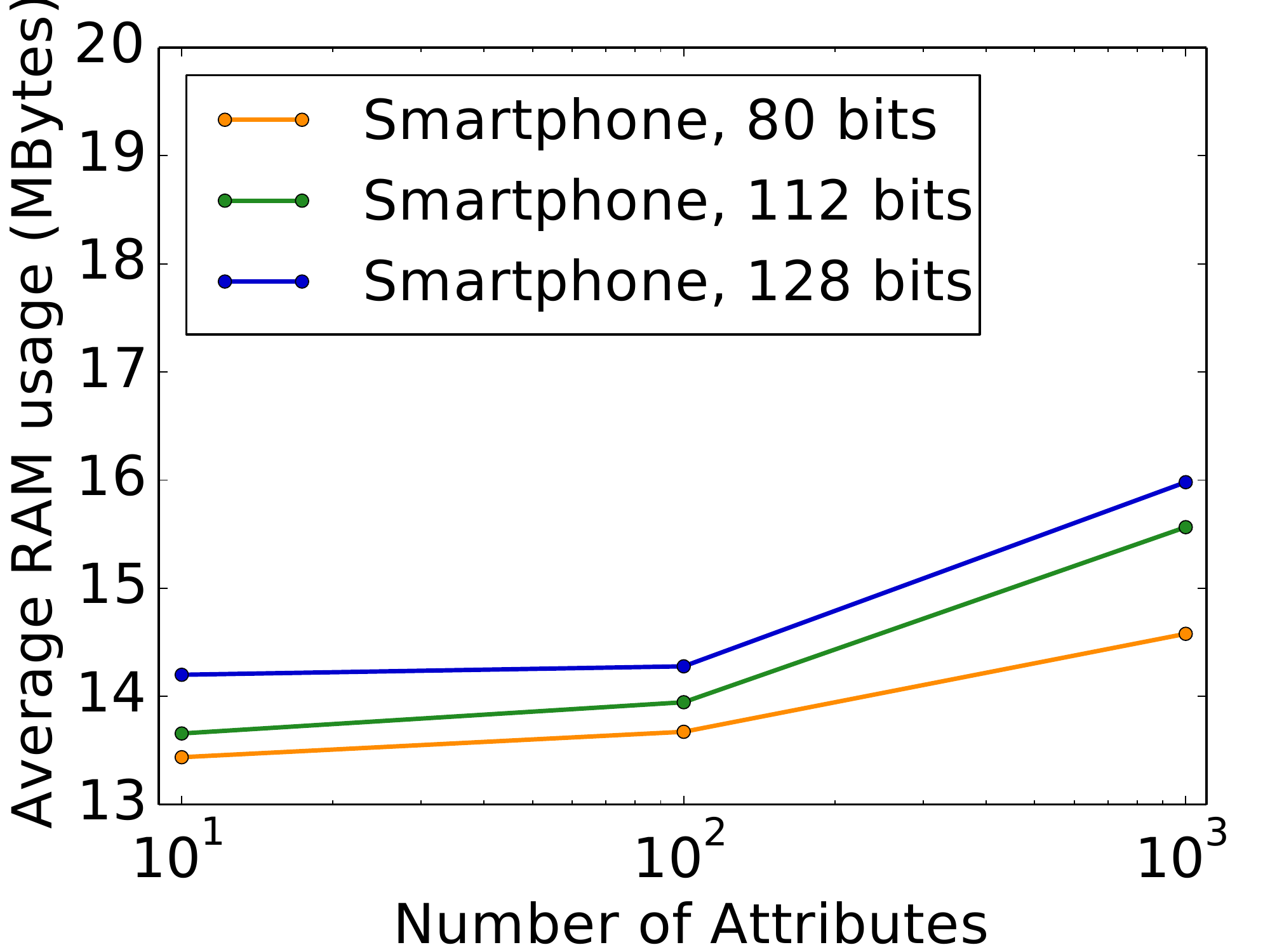}
		\caption{CP-ABE Keygen}
		\label{cpabe_keygen_RAM}
	\end{subfigure}
	\begin{subfigure}[b]{0.49\columnwidth}
		\centering
		\includegraphics[width=.95\columnwidth]{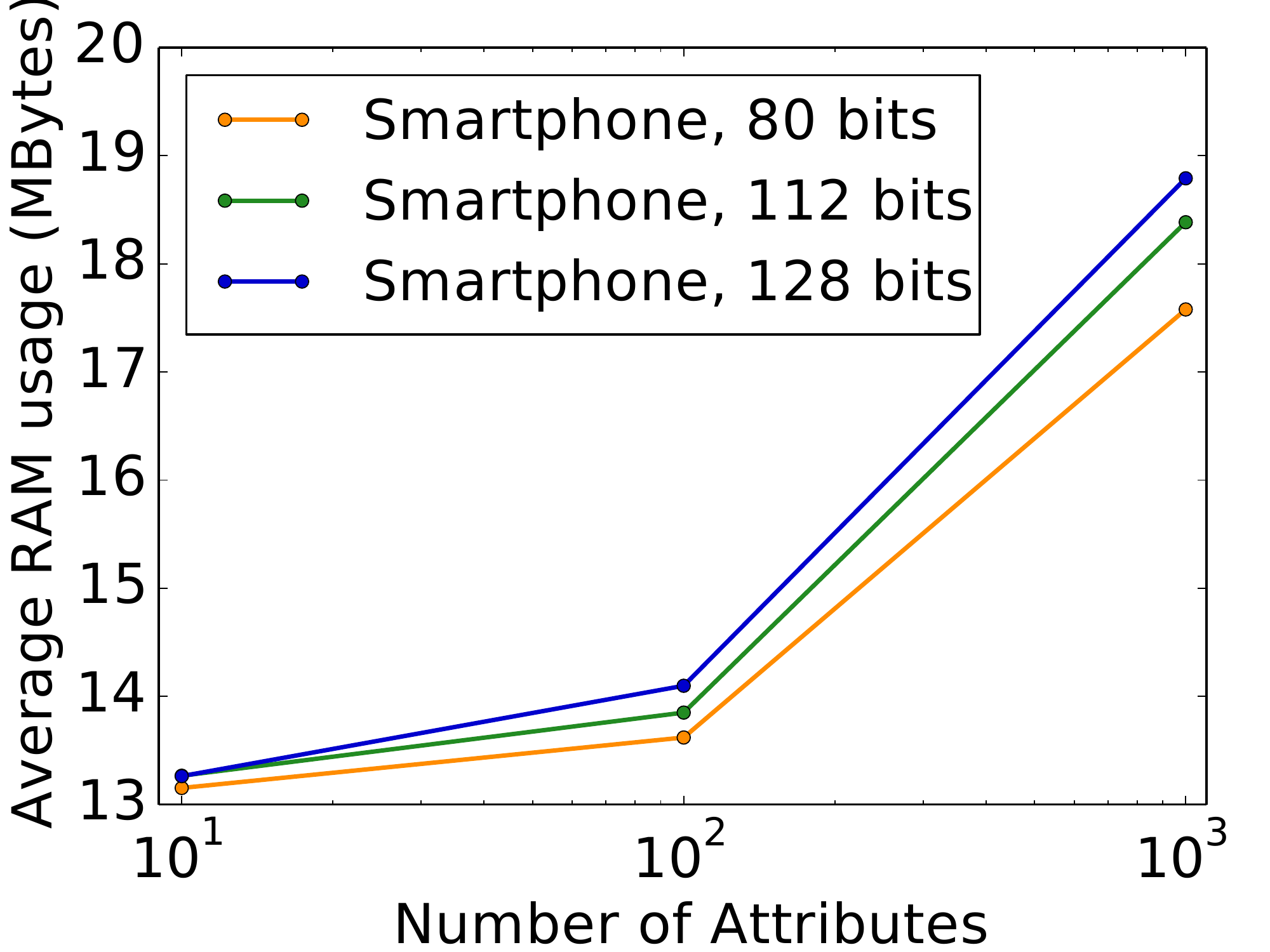}
		\caption{KP-ABE Keygen}
		\label{kpabe_keygen_RAM}
	\end{subfigure}
	\\
	\begin{subfigure}[b]{0.49\columnwidth}
		\centering
		\includegraphics[width=.95\columnwidth]{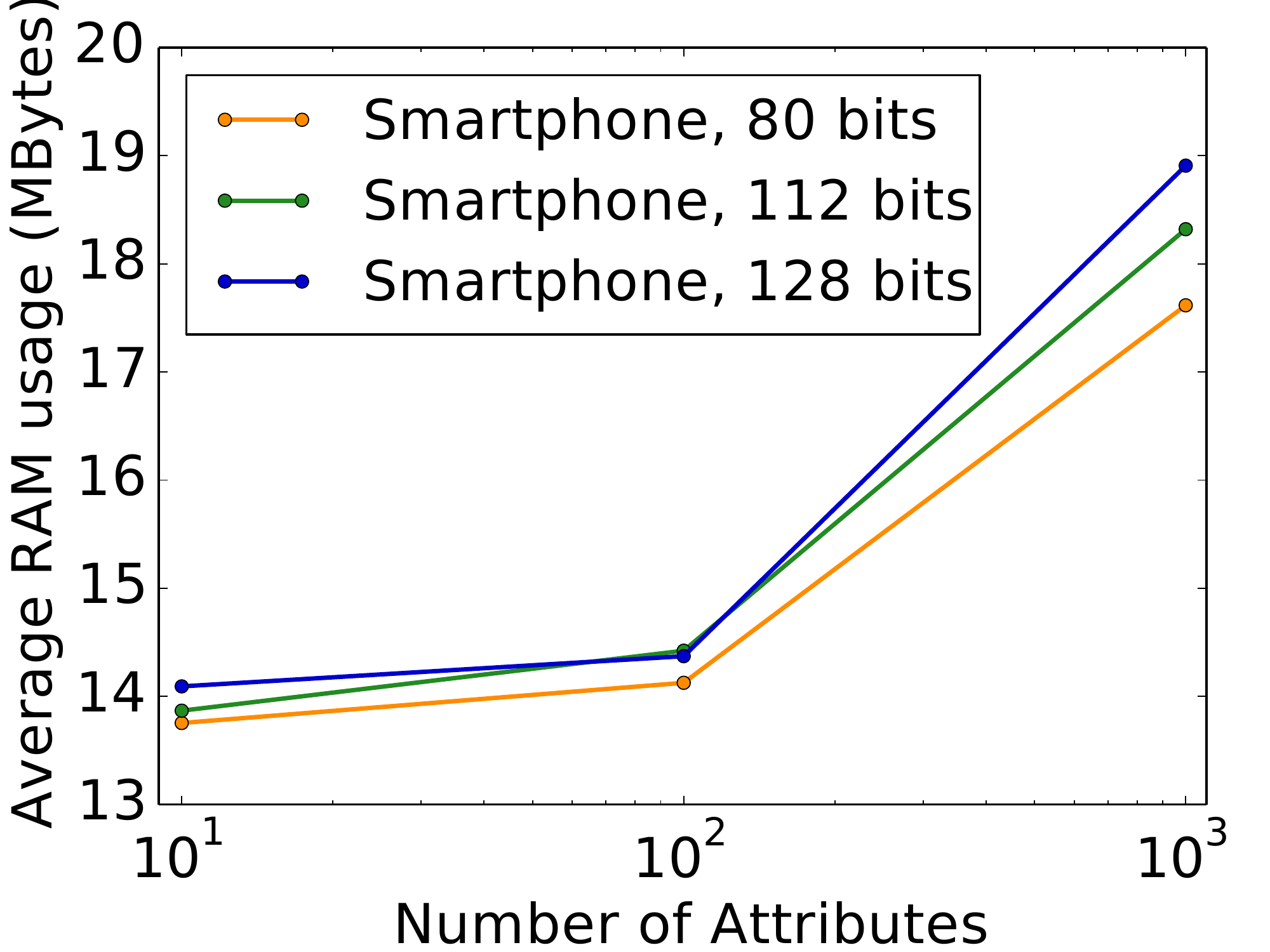}
		\caption{CP-ABE Encryption}
		\label{cpabe_enc_RAM}
	\end{subfigure}
	\begin{subfigure}[b]{0.49\columnwidth}
		\centering
		\includegraphics[width=.95\columnwidth]{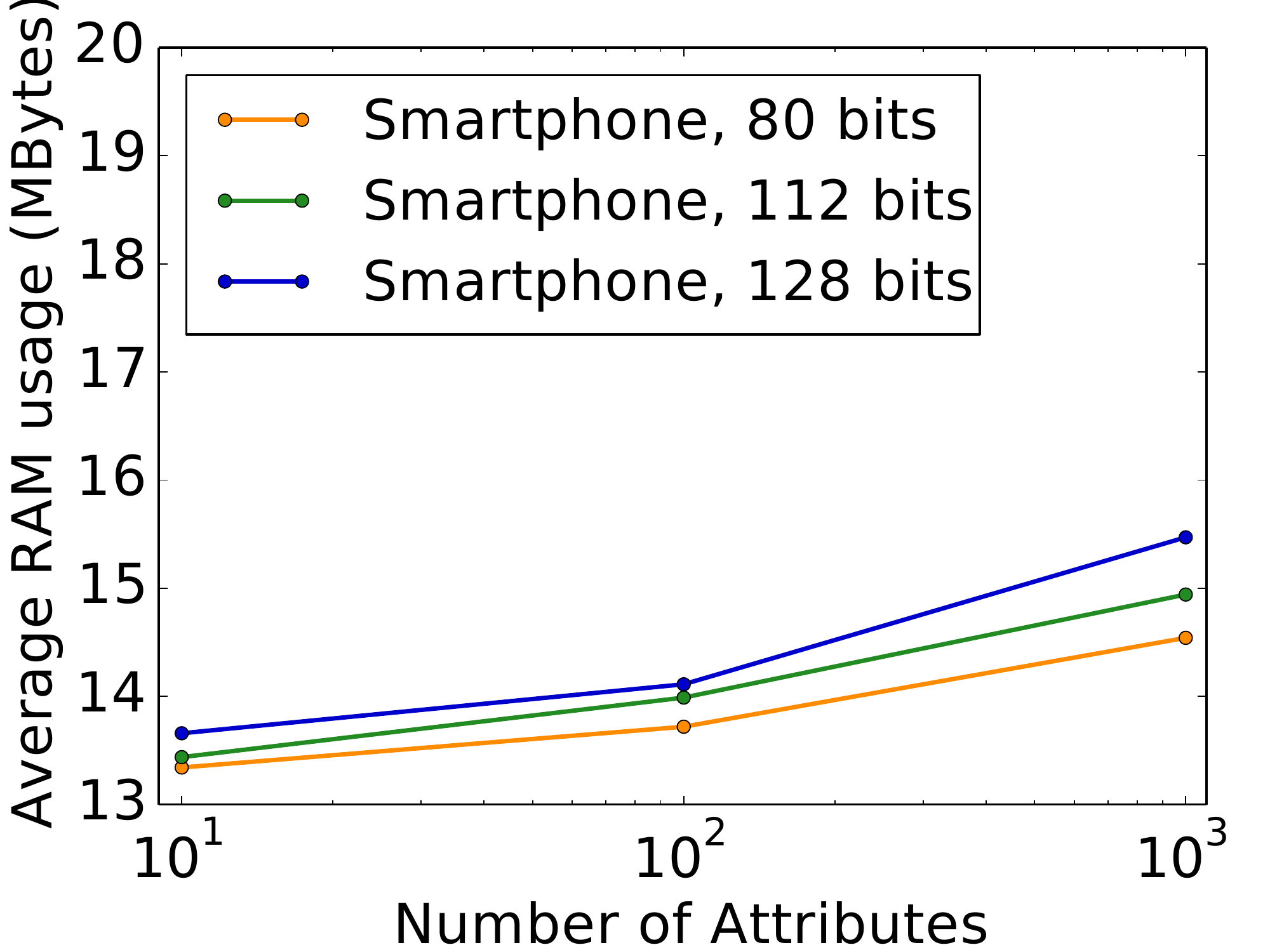}
		\caption{KP-ABE Encryption}
		\label{kpabe_enc_RAM}
	\end{subfigure}
	\\
	\begin{subfigure}[b]{0.49\columnwidth}
		\centering
		\includegraphics[width=.95\columnwidth]{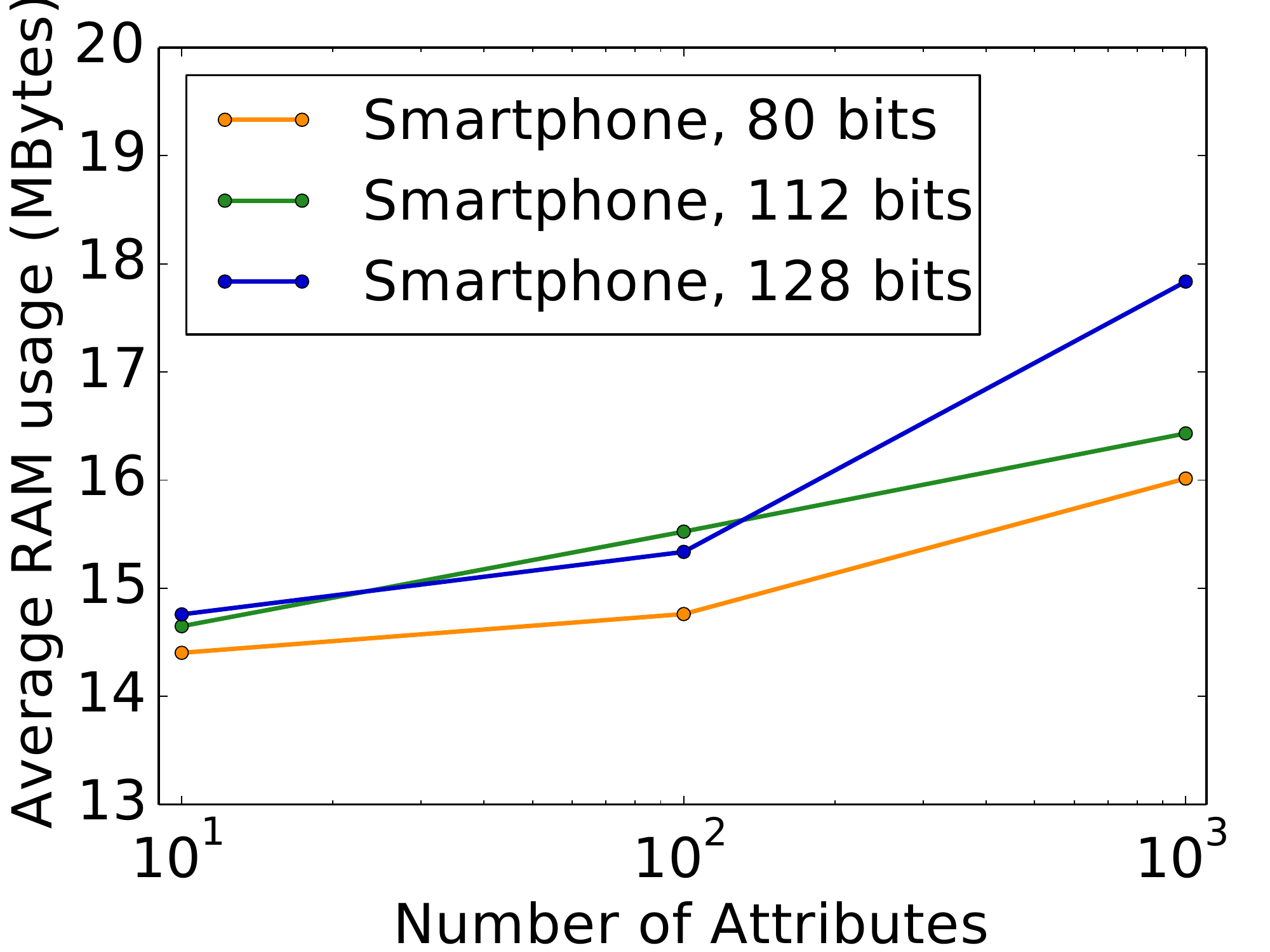}
		\caption{CP-ABE Decryption}
		\label{cpabe_dec_RAM}
	\end{subfigure}
	\begin{subfigure}[b]{0.49\columnwidth}
		\centering
		\includegraphics[width=.95\columnwidth]{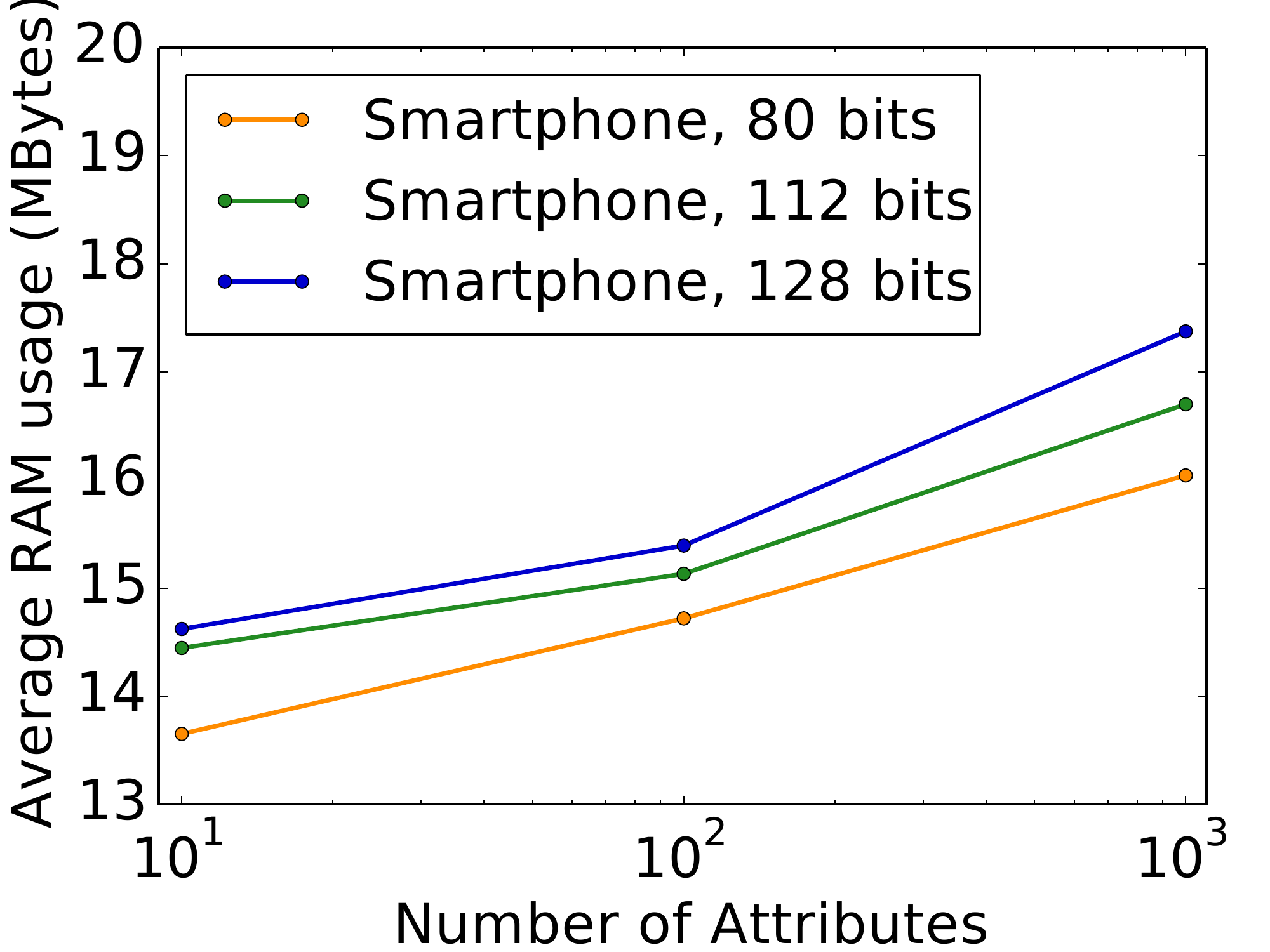}
		\caption{KP-ABE Decryption}
		\label{kpabe_dec_RAM}
	\end{subfigure}
	\caption{Average RAM usage for CP-ABE and KP-ABE on Android smartphone.}
	\label{fig:android_RAM}
\end{figure}

To better understand the behavior of \andraben~in large scale scenarios, we also measured the average memory consumption employing 10, 100 and 1000 attributes. Figure~\ref{fig:android_RAM} shows the obtained results. As expected, the amount of required RAM grows with the number of employed attributes. Here, one of the main advantages of running CP-ABE and KP-ABE on native code is the possibility to manage the heap usage directly, without relying on the Dalvik VM. Indeed, in such case the use of the expensive garbage collector significantly slows down the overall execution~\cite{6654173}.

\begin{figure}[h!]
	\centering
	\begin{subfigure}[b]{0.49\columnwidth}
		\centering
		\includegraphics[width=.95\columnwidth]{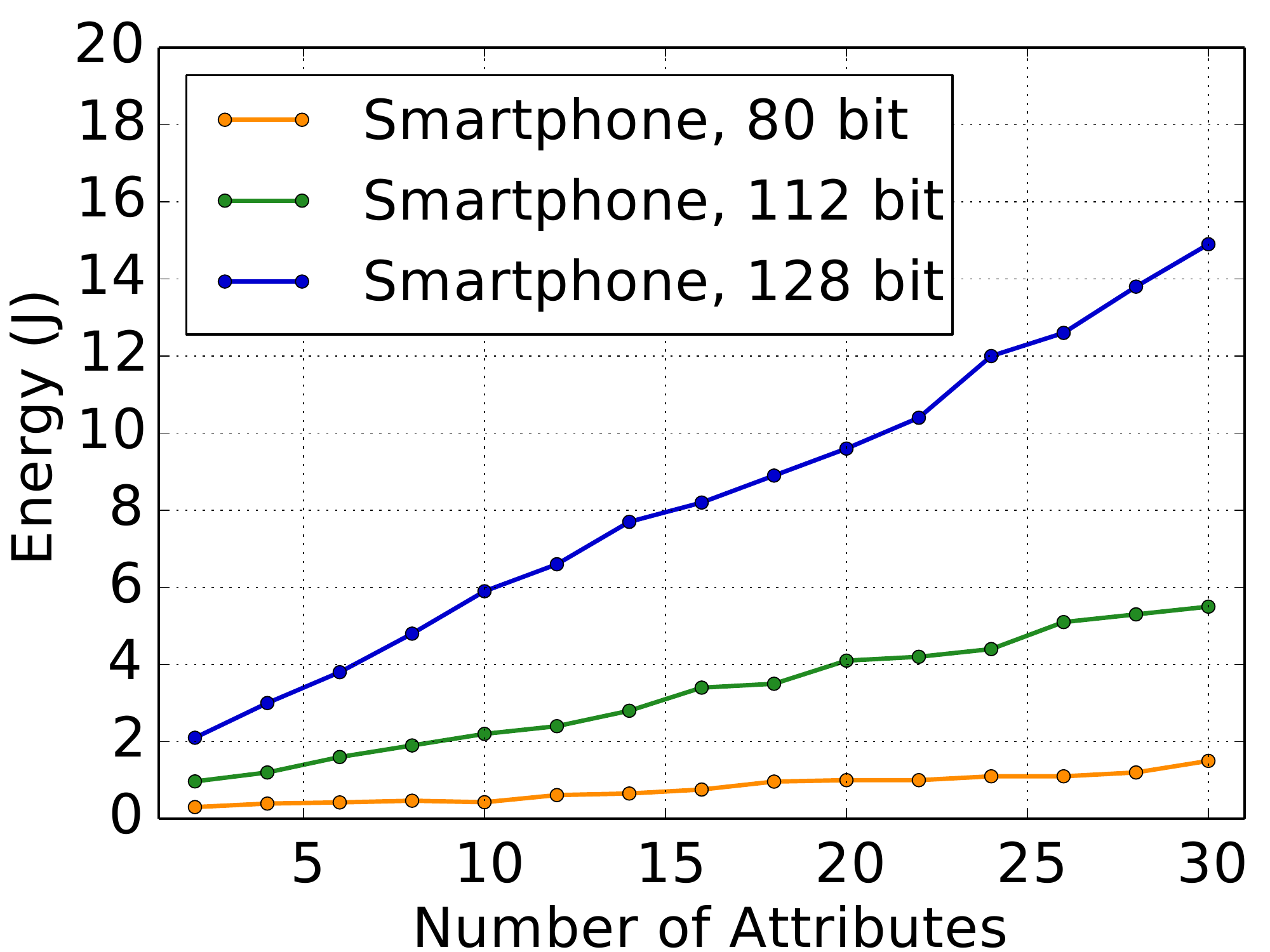}
		\caption{CP-ABE Keygen}
		\label{cpabe_keygen_android_battery}
	\end{subfigure}
	\begin{subfigure}[b]{0.49\columnwidth}
		\centering
		\includegraphics[width=.95\columnwidth]{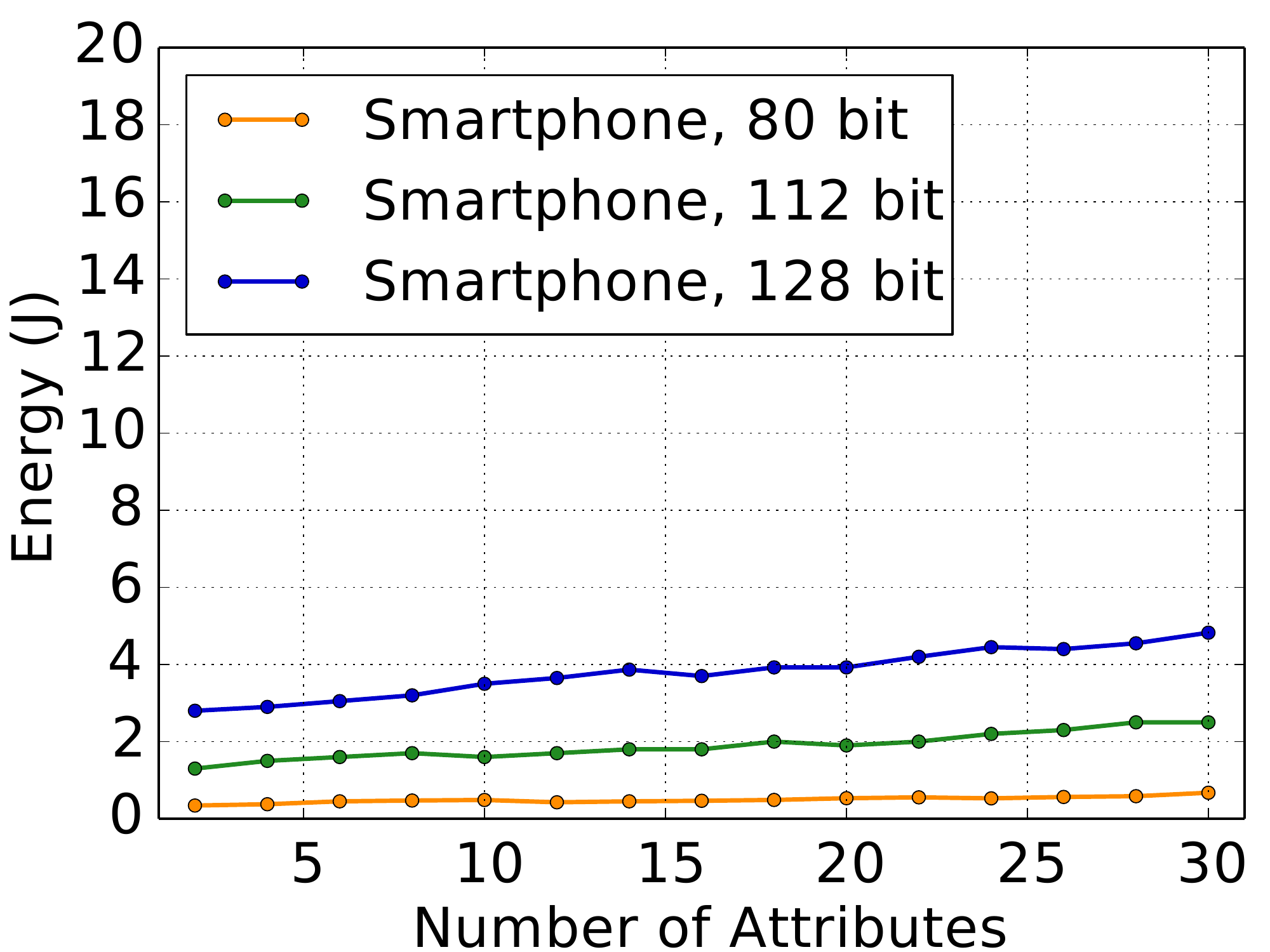}
		\caption{KP-ABE Keygen}
		\label{kpabe_keygen_android_battery}
	\end{subfigure}
	\\
	\begin{subfigure}[b]{0.49\columnwidth}
		\centering
		\includegraphics[width=.95\columnwidth]{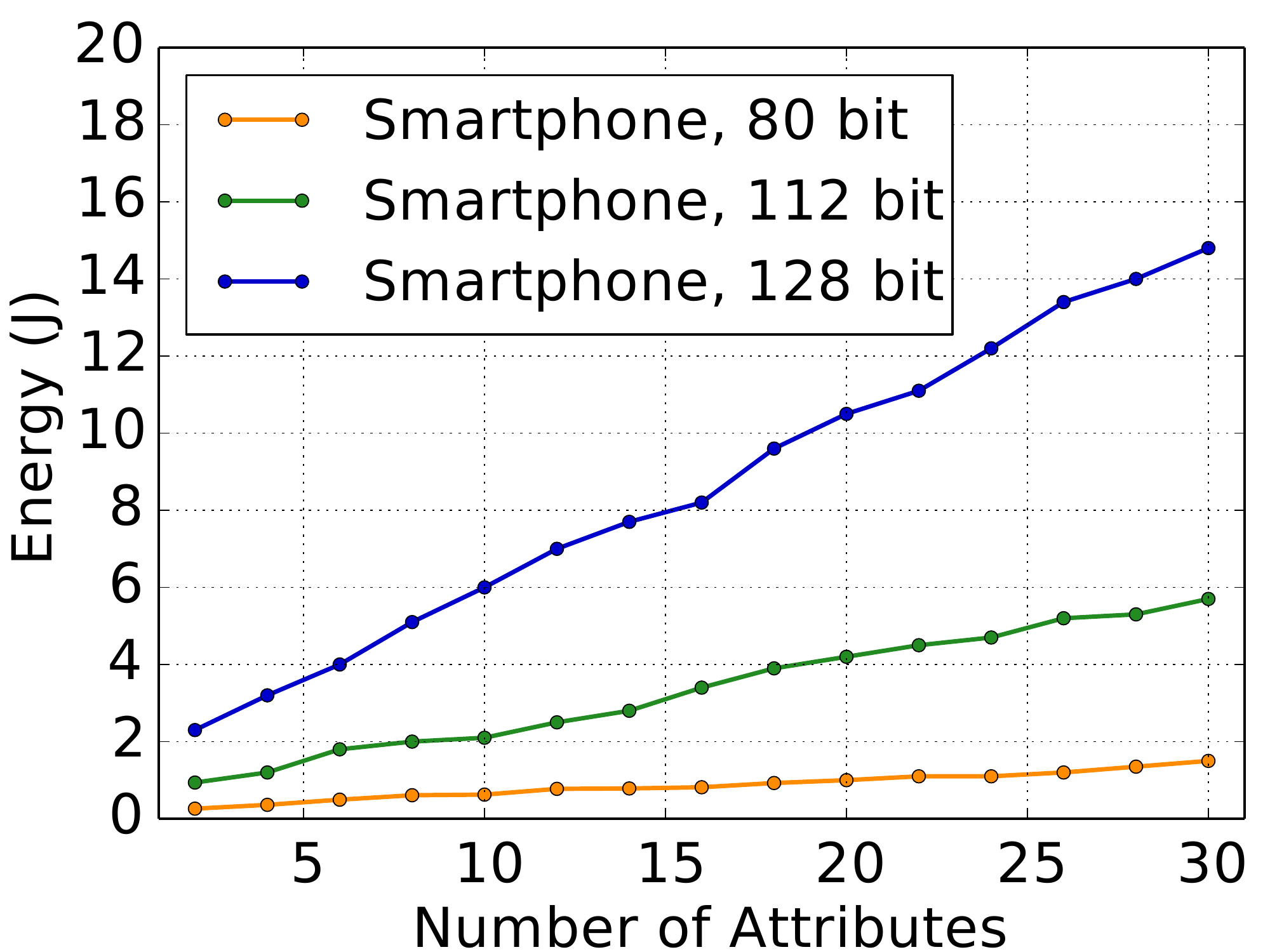}
		\caption{CP-ABE Encryption}
		\label{cpabe_enc_android}
	\end{subfigure}
	\begin{subfigure}[b]{0.49\columnwidth}
		\centering
		\includegraphics[width=.95\columnwidth]{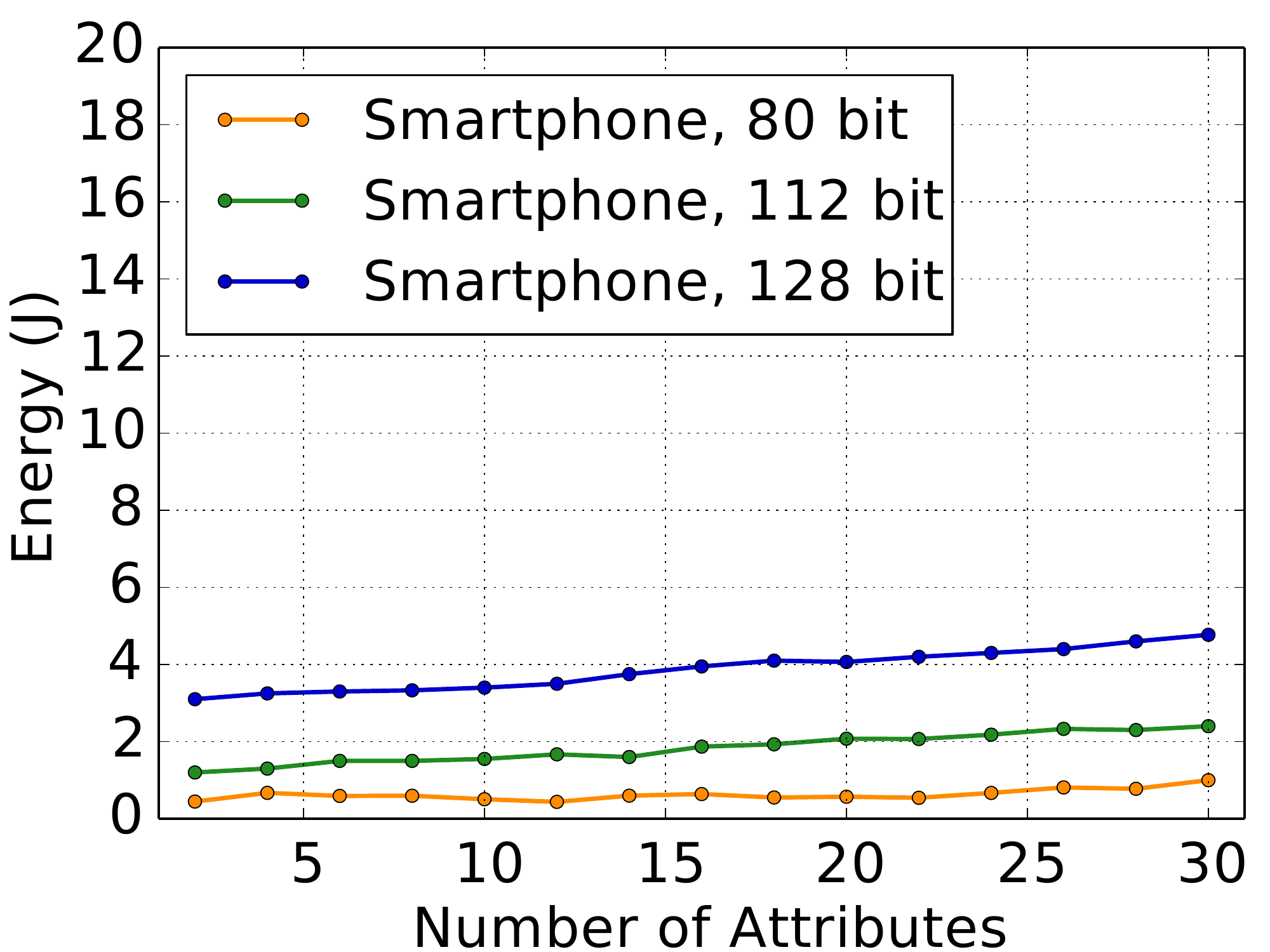}
		\caption{KP-ABE Encryption}
		\label{kpabe_enc_android_battery}
	\end{subfigure}
	\\
	\begin{subfigure}[b]{0.49\columnwidth}
		\centering
		\includegraphics[width=.95\columnwidth]{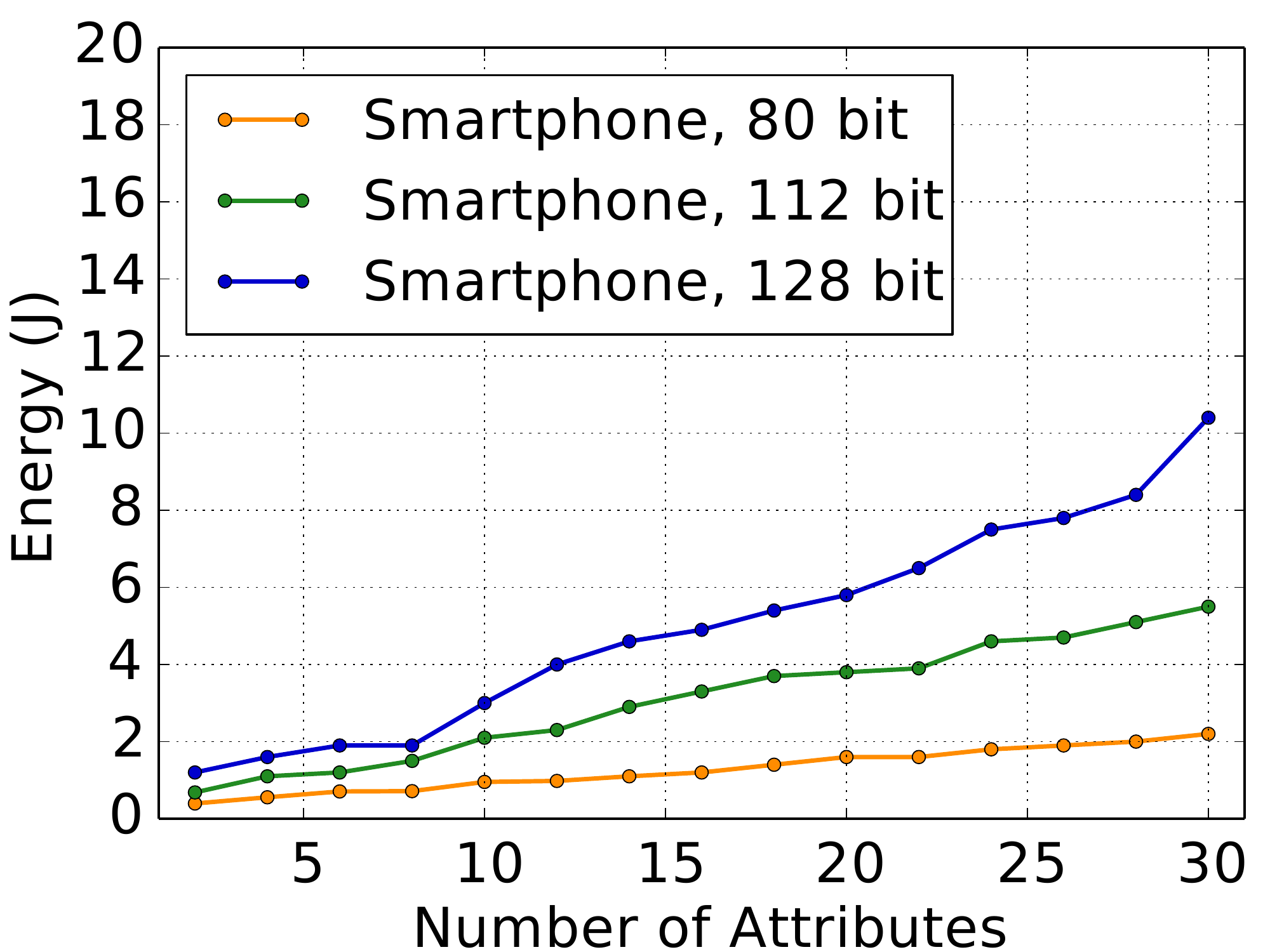}
		\caption{CP-ABE Decryption}
		\label{cpabe_dec_android}
	\end{subfigure}
	\begin{subfigure}[b]{0.49\columnwidth}
		\centering
		\includegraphics[width=.95\columnwidth]{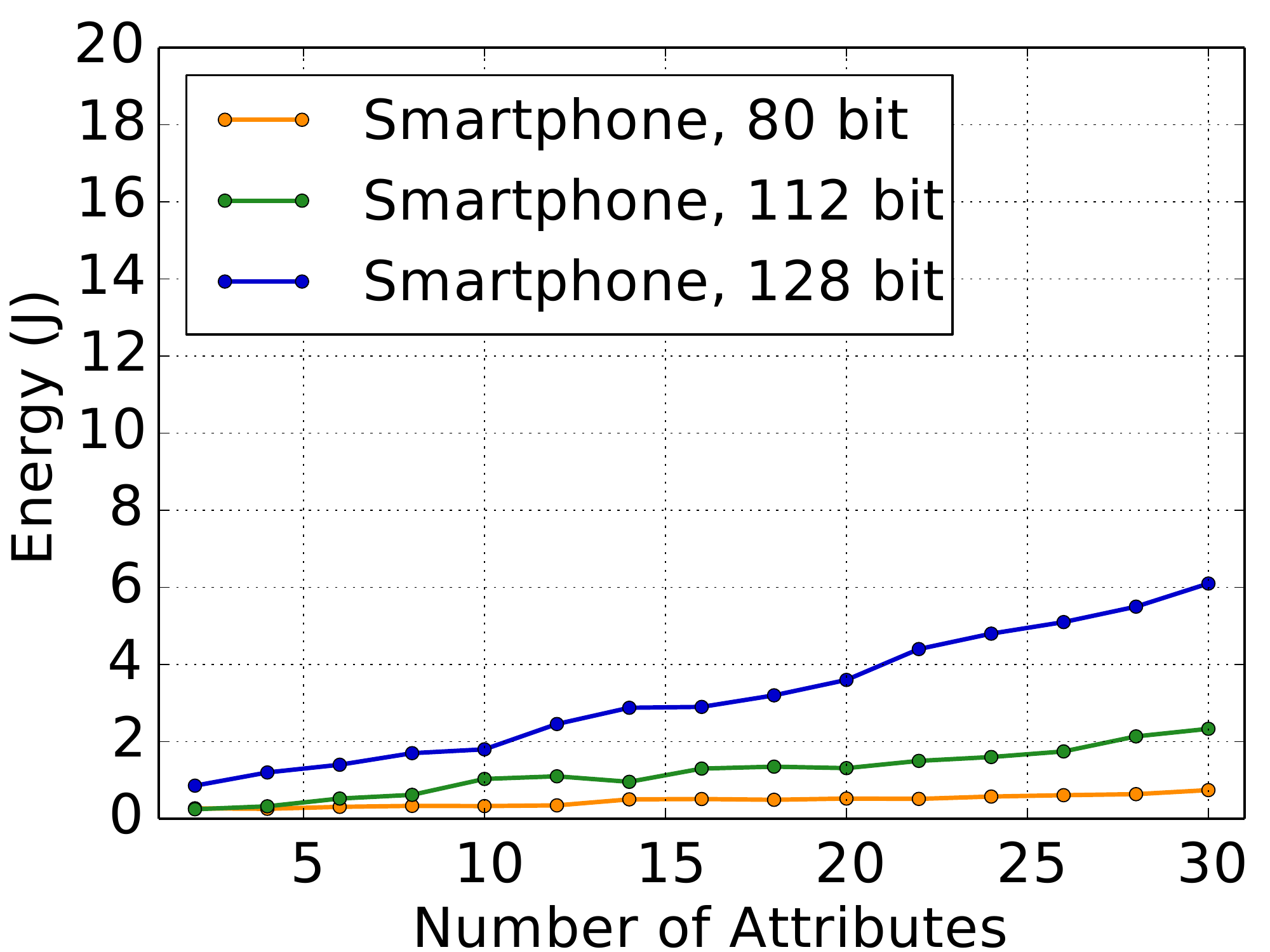}
		\caption{KP-ABE Decryption}
		\label{kpabe_dec_android_energy}
	\end{subfigure}
	\caption{Average energy consumption for CP-ABE and KP-ABE on Android smartphone.}
	\label{fig:android_battery}
\end{figure}

\paragraph{Energy Consumption} Energy consumption is a major concern in mobile devices. Therefore, a desirable implementation of a cryptographic tool for mobile devices, should consume as less energy as possible. We measured the average energy consumption for each of the CP-ABE and KP-ABE operations, by using the well-known PowerTutor Android application~\cite{powerTutor}. Figure~\ref{fig:android_battery} shows the obtained results. As we can see, with a security level of 80 and 112 bits, the energy required by both schemes remains low, making their use suitable on smartphones.

\subsection{Discussion}\label{sec:discussion}

In what follows, we provide a brief comparison of our proposal, \andraben, against the implementation proposed in~\cite{ABE_icc_2014}, on the Android platform. 

Unfortunately, the code used in~\cite{ABE_icc_2014} does not seem to be publicly available. Moreover, in~\cite{ABE_icc_2014}, the authors performed their evaluation on a device (and a specific processor in smartphone) which is not easily available anymore (an Android smartphone with a 1.60~GHz Intel Atom Z2460 processor and 1~GB RAM). For these reasons, to compare with the implementation proposed in~\cite{ABE_icc_2014}, we can only rely on the numbers reported in~\cite{ABE_icc_2014} itself. 
Most of the discussion that follows is based on this approach. Moreover, for a further validation we also re-implemented the solution proposed in~\cite{ABE_icc_2014} and performed some additional comparison.
As a side note we observe that, since our implementation is single threaded, it does not take advantage from the dual-core CPU of the device we used for our measurements. Furthermore, both the device we adopted and the one used in~\cite{ABE_icc_2014} are equipped with the same RAM memory (i.e., 1~GB).


As shown in Figure~\ref{fig:time_android}, the execution time for both CP-ABE and KP-ABE with \andraben~is significantly lower compared to the results reported in~\cite{ABE_icc_2014}. In particular, the CP-ABE key generation with \andraben~requires less than 30~s, while with the implementation proposed in~\cite{ABE_icc_2014}, key generation requires around 200~s. Similarly, \andraben~performs CP-ABE encryption and decryption in less than 30~s and 20~s, respectively, while in the implementation proposed in~\cite{ABE_icc_2014}, encryption and decryption operations take on average 70~s and 80~s, respectively. 
Moreover, the average execution time reported in~\cite{ABE_icc_2014} for all the three main KP-ABE operations, considering 26 attributes and a security level of 128 bits, is $\approx45$~s for encryption, while decryption and key generation operations require between $\approx90$~s and $\approx110$~s. Instead, our \andraben~implementation of KP-ABE requires a considerably lower execution time for each of the main operations, i.e., 12~s for decryption (Figure~\ref{kpabe_dec_android}), and 3~s for encryption (Figure~\ref{kpabe_enc_android}) and key generation (Figure~\ref{kpabe_keygen_android}).

In order to compare the energy consumption of \andraben~(which is illustrated in Figure~\ref{fig:android_battery}) with the implementation proposed in~\cite{ABE_icc_2014} for CP-ABE, let us consider 10 attributes and a security level of 128 bits. With \andraben, each of the main CP-ABE operations, i.e., key generation, encryption and decryption, consumes almost 5~J, while with the implementation proposed in~\cite{ABE_icc_2014}, key generation consumes between 70~J and 100~J, and encryption and decryption operations consume 30~J and 40~J, respectively.
Furthermore, compared to \andraben, also the KP-ABE implementation of key generation, encryption and decryption provided in~\cite{ABE_icc_2014} require a considerably higher amount of energy. Indeed, while our implementation of the three main KP-ABE operations requires less than 5~J, the implementation in~\cite{ABE_icc_2014} requires between 15~J and 40~J.

\begin{figure}[h!]
	\centering
	\begin{subfigure}[b]{0.49\columnwidth}
		\centering
		\includegraphics[width=.95\columnwidth]{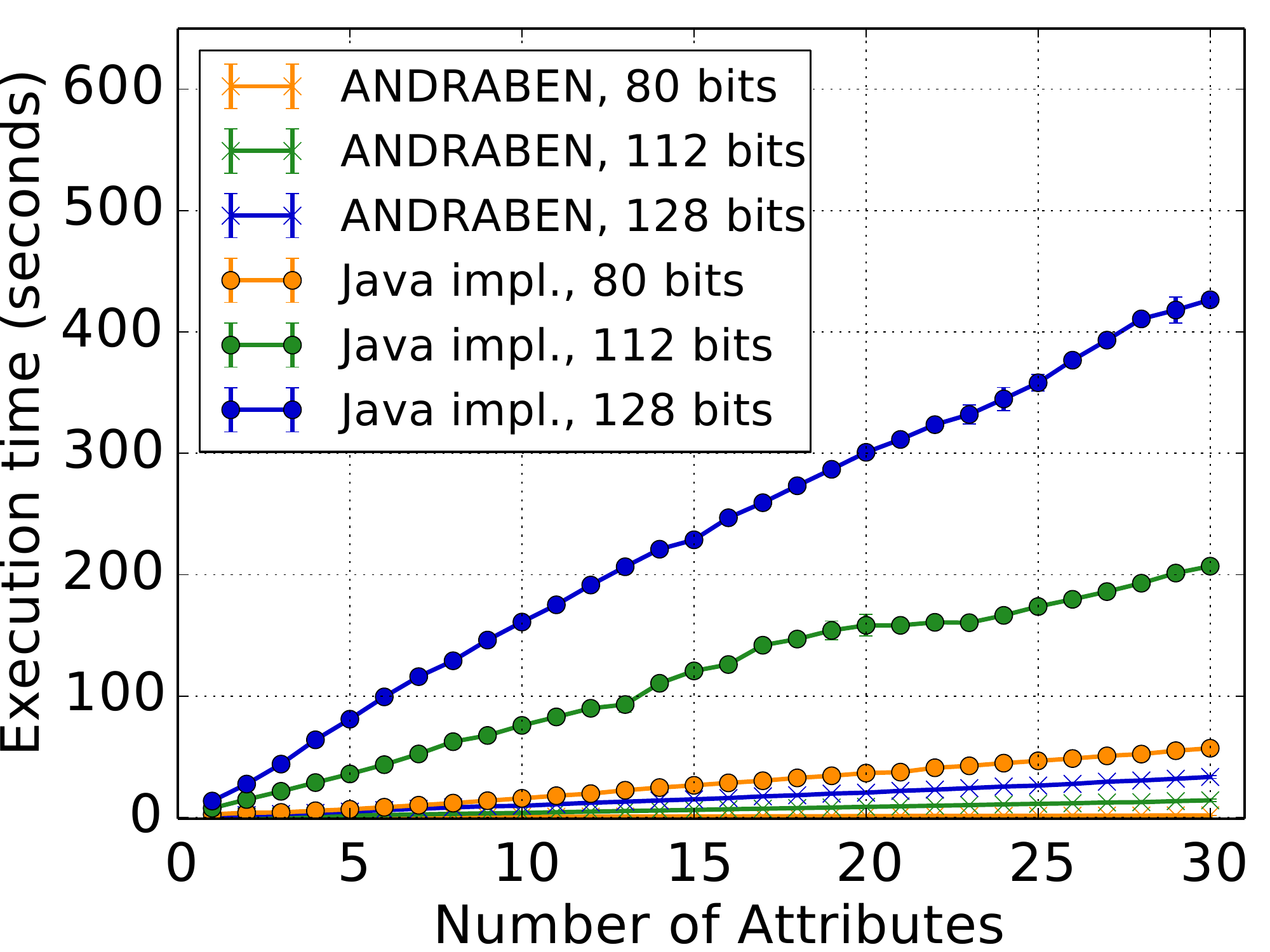}
		\caption{CP-ABE Keygen}
		\label{cpabe_keygen_android_battery}
	\end{subfigure}
	\begin{subfigure}[b]{0.49\columnwidth}
		\centering
		\includegraphics[width=.95\columnwidth]{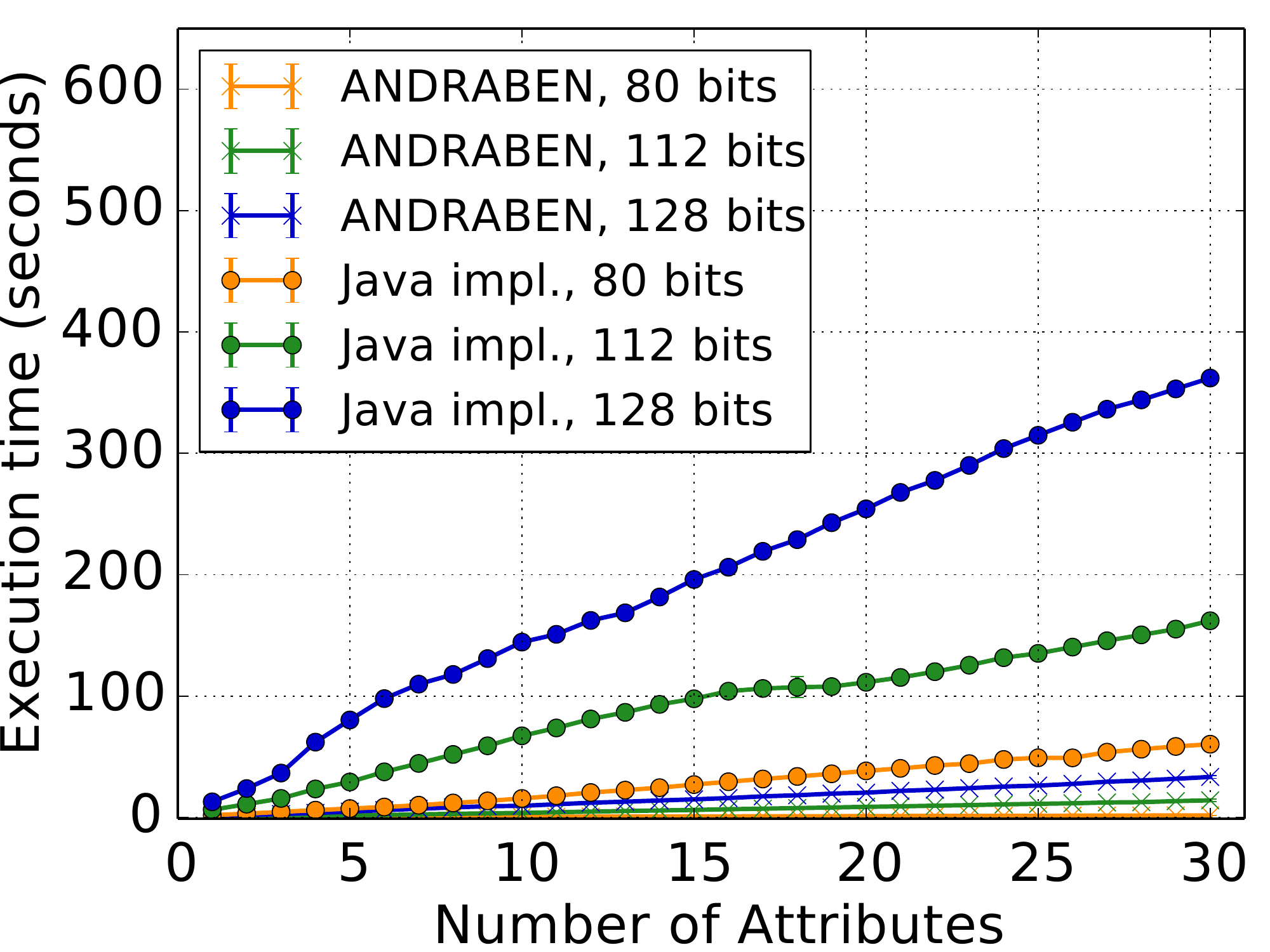}
		\caption{CP-ABE Encryption}
		\label{cpabe_enc_android}
	\end{subfigure}
	\begin{subfigure}[b]{0.49\columnwidth}
		\centering
		\includegraphics[width=.95\columnwidth]{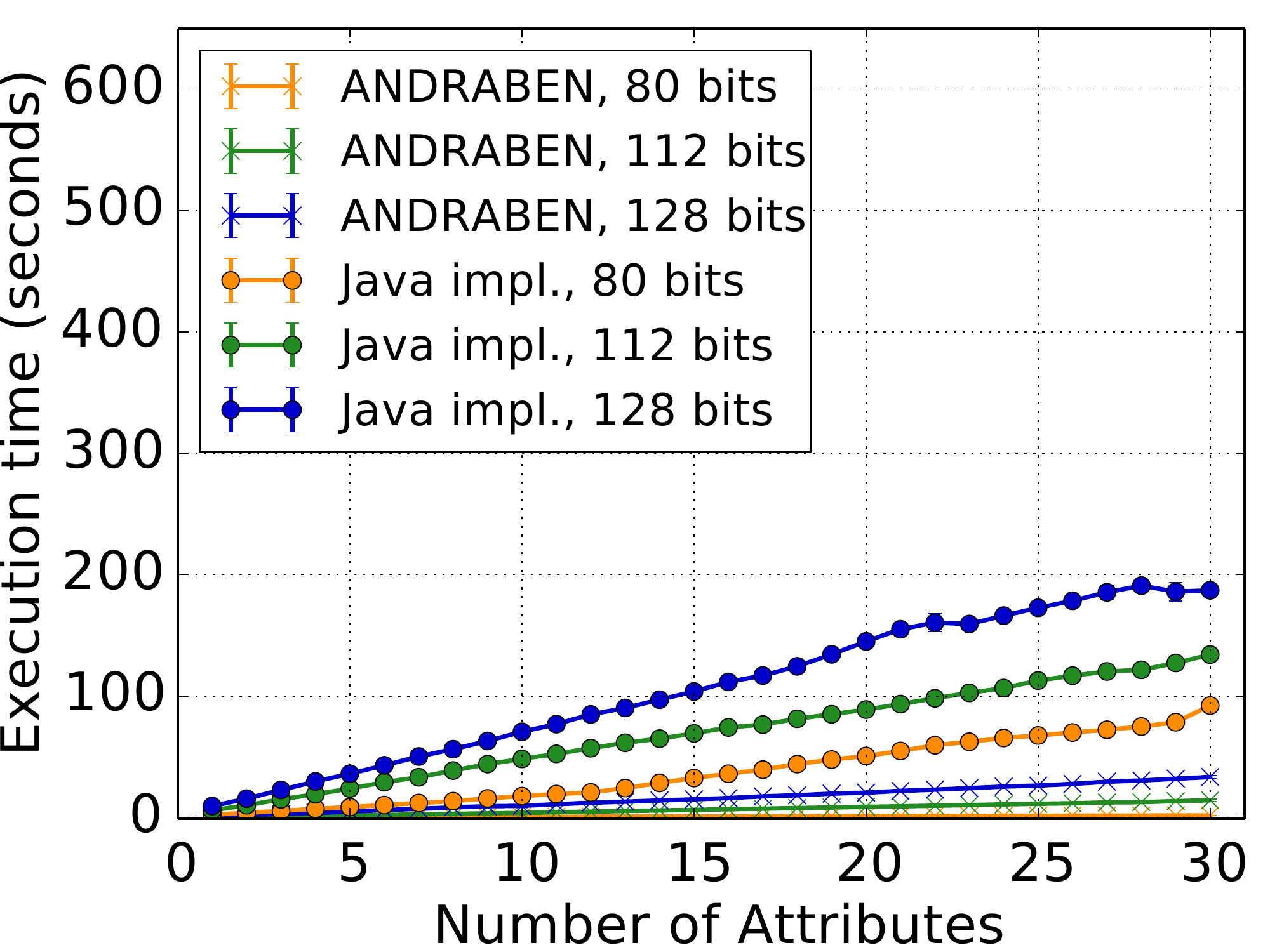}
		\caption{CP-ABE Decryption}
		\label{cpabe_dec_android}
	\end{subfigure}
	\caption{Comparison between the CP-ABE implementation of \andraben~and in~\cite{ABE_icc_2014}.}
	\label{fig:comparison}
\end{figure}

The comparison reported up to this point considered directly the performance results reported in~\cite{ABE_icc_2014}.
 However, for a further validation we also developed our own Java-based implementation of the CP-ABE scheme in~\cite{bethencourt2007ciphertext}, following the specifications reported in~\cite{ABE_icc_2014}. We evaluated our Java implementation (which we believe being similar to the one used in~\cite{ABE_icc_2014}, for which code is not available) on the same device we used to evaluate \andraben. 
Due to space limitations, we only provide a comparison of the different execution time for the two solutions, which is presented in Figure~\ref{fig:comparison}. As we can see, the execution time for CP-ABE with \andraben~is significantly lower compared to the results obtained with our Java based implementation, that in turn presents results that are consistent with the ones provided in~\cite{ABE_icc_2014}. As an example, let us consider the execution time obtained with 25 attributes and a security level of 128 bits. Our CP-ABE \andraben~implementation requires $\approx$25~s to perform the key generation, while the Java-based implementation requires $\approx$360~s to perform the same task. Similarly, the CP-ABE encryption and decryption operations with \andraben~require on average $\approx$26~s and $\approx$19~s, while our Java-based implementation performs encryption and decryption in $\approx$340~s and $\approx$172~s, respectively. 

Overall, we can conclude that, compared with the approach discussed in~\cite{ABE_icc_2014}, \andraben~provides significantly better performance, in terms of execution time, memory and CPU usage, and energy consumption.

\section{Conclusion}\label{sec:conclusion} 
With the increasing use of cloud environment and smart devices connected to the Internet of Things, exchanged data confidentiality and access control to the stored data become a challenging issue. Attribute-Based Encryption is one of the best solutions that can be used to satisfy users privacy concerns~\cite{ABE_icc_2014}. However, its performance on resource constraint devices is a challenging issue, and still represents a big concern for researchers willing to use ABE to develop novel privacy-preserving and access control solutions for such devices.

In this paper, we studied the feasibility of applying ABE on smartphone devices and presented \andraben, an implementation of ABE in C language.
We also provided a comparative analysis with a similar research study~\cite{ABE_icc_2014} in which the authors proposed a Java-based implementation of ABE for Android smartphone.
Based on the results of our thorough experiments, we conclude that using ABE on Android smartphones and similar devices is feasible. 
The evidence that we bring in this paper will be a reference for applicability of ABE in resource-constrained devices.

\balance
\bibliographystyle{abbrv}
\bibliography{ref}

\end{document}